\documentclass{aa}
\usepackage{graphicx}

\def\mb#1{\setbox0=\hbox{$#1$}\kern-.025em\copy0\kern-\wd0
\kern-0.05em\copy0\kern-\wd0\kern-.025em\raise.0233em\box0}

\newcommand{\eq}[1]{(\ref{#1})}

\begin{document}

   \title{Statistical mechanics and phase diagrams of rotating self-gravitating fermions }

\titlerunning{Rotating self-gravitating fermions}

   \author{P.H. Chavanis\inst{1} \and M. Rieutord \inst{2,3}}

\institute{ Laboratoire de Physique Th\'eorique, Universit\'e Paul
Sabatier, 118 route de Narbonne 31062 Toulouse, France\\
\email{chavanis@irsamc.ups-tlse.fr} \and Observatoire
Midi-Pyr\'en\'ees, 14 av. E. Belin, F-31400 Toulouse, France
\and Institut Universitaire de France
\\ \email{rieutord@ast.obs-mip.fr}}

   \date{\today}

   \abstract{We compute statistical equilibrium states of rotating
   self-gravitating fermions by maximizing the Fermi-Dirac entropy at
   fixed mass, energy and angular momentum. We describe the phase
   transition from a gaseous phase to a condensed phase (corresponding
   to white dwarfs, neutron stars or fermion balls in dark matter
   models) as we vary energy and temperature.  We increase the
   rotation up to the Keplerian limit and describe the flattening of
   the configuration until mass shedding occurs. At the maximum
   rotation, the system develops a {\it cusp} at the equator.  We draw
   the equilibrium phase diagram of the rotating self-gravitating
   Fermi gas and discuss the structure of the caloric curve as a
   function of degeneracy parameter (or system size) and angular
   velocity. We argue that systems described by the Fermi-Dirac
   distribution in phase space do not bifurcate to non-axisymmetric
   structures when rotation is increased, in continuity with the case
   of polytropes with index $n>0.808$ (the Fermi gas at $T=0$
   corresponds to $n=3/2$). This differs from the study of Votyakov
   et al. (2002) who consider a Fermi-Dirac distribution in
   configuration space appropriate to stellar formation and find
   ``double star'' structures (their model at $T=0$ corresponds to
   $n=0$). We also consider the case of classical objects described by
   the Boltzmann entropy and discuss the influence of rotation on the
   onset of gravothermal catastrophe (for globular clusters) and
   isothermal collapse (for molecular clouds). On general grounds, we
   complete previous investigations concerning the nature of phase
   transitions in self-gravitating systems. We emphasize the
   inequivalence of statistical ensembles regarding the formation of
   binaries (or low-mass condensates) in the microcanonical ensemble
   (MCE) and Dirac peaks (or massive condensates) in the canonical
   ensemble (CE). We also describe an {\it hysteretic cycle} between
   the gaseous phase and the condensed phase that are connected by a
   ``collapse'' or an ``explosion''.

  \keywords{Stellar
   dynamics-hydrodynamics, instabilities } }

   \maketitle


\section{Introduction}
\label{sec_introduction}

It is striking to observe that self-gravitating systems follow a kind
of organization despite the diversity of their initial conditions and
their environment. This organization is illustrated by morphological
classification schemes such as the Hubble sequence for galaxies and by
simple rules which govern the structure of individual self-gravitating
systems. For example, elliptical galaxies display a quasi-universal
luminosity profile described by de Vaucouleur's $R^{1/4}$ law and most
of globular clusters are well-fitted by the Michie-King model. On the
other hand, the rotation curves of spiral galaxies appear to be flat
and this striking observation can be explained by the presence of a
dark matter halo with a density profile decreasing as $r^{-2}$ at
large distances (Binney \& Tremaine 1987).

The question that naturally emerges is what determines the particular
configuration to which a self-gravitating system settles. It is
possible that their actual configuration crucially depends on the
conditions that prevail at their birth and on the details of their
evolution. However, in view of their apparent regularity, it is
tempting to investigate whether their organization can be favoured by
some fundamental physical principles like those of thermodynamics and
statistical physics. We ask therefore if the actual states of
self-gravitating systems are not simply more probable than any other
possible configuration, i.e. if they cannot be considered as {\it
maximum entropy states} (Chavanis 2002a).

This thermodynamical approach may be particularly relevant for
globular clusters and elliptical galaxies whose distribution function
is close to the isothermal one (at least in the inner region of the
system). In the case of globular clusters, the relaxation proceeds via
two-body encounters and this ``collisional'' evolution is governed by
the Fokker-Planck-Landau kinetic equation for which a H-theorem is
available (see Kandrup 1981 for a precise discussion). This
collisional evolution can lead to the establishment of a statistical
equilibrium state. This equilibrium state is usually described by the
Michie-King model, a truncated isothermal, which takes into account
the escape of high energy stars due to tidal forces (Michie 1963, King
1966).  In fact, such equilibria are only {\it metastable} (Antonov
1962) and the evolution continues for longer times with the formation
of binaries (H\'enon 1961). By contrast, for elliptical galaxies,
two-body encounters are completely negligible (the corresponding
relaxation time $t_{{\it coll}}$ exceeds the age of the universe by
many orders of magnitude) and the galaxy dynamics is described by the
{Vlasov equation}, i.e. collisionless Boltzmann equation. Yet,
collisionless stellar systems can reach a metaequilibrium state as a
result of a {\it violent relaxation} (Lynden-Bell 1967). The
statistical prediction of this metaequilibrium state is complicated
and depends on the initial conditions (Lynden-Bell 1967, Chavanis
2003a). Furthermore, the metaequilibrium state actually reached by the
system may differ from the statistical prediction due to {\it
incomplete relaxation}. Therefore, $H$-functions
must be constructed to account for the structure of elliptical
galaxies (Tremaine, H\'enon \& Lynden-Bell 1987, Hjorth \& Madsen
1993, Chavanis 2003a,c).

The thermodynamics of self-gravitating systems started with the
discoveries by Emden (1907), Bonnor (1956) and Antonov (1962) that a
self-gravitating isothermal gas can be in hydrostatic equilibrium only
in a limited range of thermodynamical parameters (energy, temperature 
and external pressure). Outside this range, the thermodynamical
potential (entropy, free energy or Gibbs energy) has no extremum and
the system is expected to collapse under its own gravity (Lynden-Bell
\& Wood 1968). A ``core-halo'' phase halting the
collapse can be evidenced if we introduce appropriate small-scale
cut-offs. Different regularizations have been proposed such as quantum
degeneracy, hard spheres, soften potential,... Then, the shape of the
caloric curve $T(E)$ depends on an additional parameter $\mu$ related
to the inverse of the small-scale cut-off (see Chavanis 2002b). For
$\mu\rightarrow +\infty$, one recovers the classical results of Emden,
Bonnor and Antonov. As the cut-off increases, it is possible to
describe phase transitions between a ``gaseous'' phase (independent on
the small-scale cut-off) and a ``condensed'' phase with a core-halo
structure. It is important to stress that statistical ensembles are
not interchangeable for self-gravitating systems (Thirring 1970,
Lynden-Bell \& Lynden-Bell 1977, Padmanabhan 1990) and that phase
transitions occur not only in the canonical ensemble but also in the
{\it microcanonical} ensemble if the cut-off is sufficiently small
(Stahl {et al.} 1994, Chavanis 2002b, Chavanis \& Ispolatov
2002). Microcanonical phase transitions are a rather new phenomenon
associated with the existence of metastable states (local entropy
maxima).

In the previous studies, the system is supposed to be
non-rotating. The thermodynamics of rotating stellar systems was
studied by Lagoute \& Longaretti (1996) in the framework of
Michie-King models describing globular clusters. The case of a
{slowly} rotating isothermal gas confined within a spherical box was
treated by Chavanis (2002c) using analytical methods introduced by
Milne (1923) and Chandrasekhar (1933) for distorted polytropes.
Recently, Fliegans \& Gross (2002) and Votyakov {et al.} (2002)
have considered the statistical mechanics of a {rapidly} rotating
self-gravitating gas in relation with the process of star formation.
They showed that phase transitions from axisymmetric to
non-axisymmetric structures (``double stars'') occur at sufficiently
high values of angular momentum. These authors provide an interesting
phase diagram of rotating self-gravitating systems for arbitrary
values of angular momentum and energy. They treat the particles as
hard spheres (of size $a$) and consider a relatively {\it large} value of
the small-scale cut-off.  Fliegans \& Gross (2002) evaluate the
density of state $g(E,L)$ numerically for a finite number $N$ of
particles in two-dimensions.  Votyakov {et al.} (2002) use a
mean-field approximation $N\gg 1$ and adopt a distribution
$\rho(\Phi)$ corresponding to a Fermi-Dirac statistics in {\it
configuration} space. This model is expected to take into account hard
sphere effects, where the hard spheres are introduced heuristically
to fix an upper limit to the local density. 

In this paper, we shall study the structure of rotating
self-gravitating systems described by a Fermi-Dirac distribution
function $f=f(\epsilon)$ in {\it phase} space.  This is the usual
distribution function of quantum particles (fermions) due to Pauli's
exclusion principle. This distribution function describes electrons in
white dwarf stars (Chandrasekhar 1942), neutrons in neutron stars
(Hertel \& Thirring 1971) and massive neutrinos in dark matter models
(Bilic \& Viollier 1997). Incidentally, this distribution function 
also occurs in the context of the violent relaxation of
collisionless stellar systems (Chavanis
\& Sommeria 1998). In the latter, the statistical approach refers to a
{\it coarse-graining} of the distribution function and the effective
exclusion principle is a consequence of the Liouville theorem
associated with the collisionless nature of the evolution (Lynden-Bell
1967, Chavanis 2002d).

The distinction between the distribution used by Votyakov et
al. (2002) and the ordinary Fermi-Dirac distribution function in phase
space is crucial, regarding the structure of the condensed phase, when
the system is rotating. At $T=0$ (or $E=E_{min}$), the former leads to
a {\it homogeneous} body $(\rho={\rm Cst.})$ while the latter leads to
a polytrope with index $n=3/2$ corresponding to a classical white
dwarf star ($p=K\rho^{5/3}$). Now, it is well-known that a uniformly
rotating homogeneous mass is axisymmetric (spheroidal) for slow
rotations (Maclaurin sequence) but bifurcates to non-axisymmetric
configurations at sufficiently high rotations (Chandrasekhar
1969). These non-axisymmetric structures are usually ellipsoidal
(Jacobi sequence) but more complex structures (pear-shaped,
dumbbell,...) have also been found.  The fission of the
self-gravitating fluid into a binary state is also possible (Hachisu
\& Eriguchi 1984) and these binary states appear in the study of
Votyakov et al. (2002) at low temperatures and high rotations. On
general grounds, the fission instability has been extensively
discussed in relation with the formation of binary stars and the
formation of the moon (in the non-symmetrical case). However, the
fission scenario may not be the relevant mechanism and it is almost
abandoned at present. Binary stars are more likely to result from the
fragmentation of the molecular cloud (Boss 1988) and the moon has
probably been detached from the protoearth after a collision with a
Mars-sized body (Boss 1986). In contrast to incompressible fluid
masses, a polytropic configuration with index $n>0.808$ does {\it not}
bifurcate to non-axisymmetric structures when angular velocity is
increased; it remains axisymmetric until the Keplerian limit is
reached (James 1964). At that point there is mass shedding at the
equator (with the formation of a cusp) and disruption of the system.
This concerns in particular polytropes of index $n=3/2$ (white dwarfs)
which are the zero temperature limit of the Fermi distribution (Fowler
1926). This picture probably persists at finite temperature where we
have, typically, a Fermi condensate surrounded by a
``vapour''. Therefore, the phase diagram corresponding to the
Fermi-Dirac statistics in phase space is expected to differ from the
one calculated by Votyakov et al. (2002). In this paper, we propose to
compute this phase diagram for arbitrary rotation and degeneracy
parameter (or system size) in both microcanonical and canonical
ensembles.  We shall restrict ourselves to axisymmetric single-cluster
configurations because our code is limited to such
configurations. According to James' result, if other structures exist
(double-clusters, rings,...), they should bifurcate {\it
discontinuously} from the spheroidal sequence as rotation
increases. Such bifurcations will not be considered in this paper.
The present study completes the description of non-rotating or slowly
rotating fermionic structures discussed by Chavanis (2002b,c).

\section{Statistical mechanics of rotating self-gravitating systems}
\label{sec_smrot}

\subsection{The maximum entropy state}
\label{sec_mf}

Consider a system of $N$ particles, each of mass $m$, interacting via
Newtonian gravity. We allow the system to have a non-vanishing angular
momentum. Let $f({\bf r},{\bf v},t)$ denote the distribution function
of the system, i.e. $f({\bf r},{\bf v},t)d^{3}{\bf r}d^{3}{\bf v}$
gives the mass of particles whose position and velocity are in the
cell $({\bf r},{\bf v};{\bf r}+d^{3}{\bf r},{\bf v}+d^{3}{\bf v})$ at
time $t$. The integral of $f$ over the velocity determines the spatial
density
\begin{equation}
\rho=\int fd^{3}{\bf v}.
\label{mf1}
\end{equation}
The mass and angular momentum of the configuration are given by
\begin{equation}
M=\int \rho d^{3}{\bf r},
\label{mf2}
\end{equation}
\begin{equation}
{\bf L}=\int f\ {\bf r}{\times} {\bf v} \ d^{3}{\bf r}d^{3}{\bf v}.
\label{mf3}
\end{equation}
On the other hand, in the mean-field approximation, the energy can be expressed
as
\begin{equation}
E={1\over 2}\int f v^{2}d^{3}{\bf r}d^{3}{\bf v}+{1\over 2}\int \rho\Phi d^{3}{\bf r}=K+W,
\label{mf4}
\end{equation}
where $K$ is the kinetic energy and $W$ the potential energy. The gravitational potential $\Phi$ is related to the density by the Newton-Poisson equation
\begin{equation}
\Delta\Phi=4\pi G\rho.
\label{mf5}
\end{equation}

Let us consider different systems of astrophysical interest that can
be approached by statistical mechanics. (i) The equilibrium states of
collisional stellar systems (e.g., globular clusters), reached for
$t>t_{relax}\sim {N\over \ln N} t_D$ (where $t_{D}$ is the dynamical
time), are obtained by maximizing the Boltzmann entropy (Antonov 1962,
Ogorodnikov 1965, Lynden-Bell \& Wood 1968)
\begin{equation}
S=-\int f\ln f d^{3}{\bf r}d^{3}{\bf v},
\label{mf6b}
\end{equation}
at fixed mass $M$, energy $E$ and angular momentum ${\bf L}$. This is
a condition of thermodynamical stability for the Hamiltonian N-stars
system. For globular systems, only the microcanonical ensemble makes sense
and equilibrium states exist only above the Antonov
energy. Furthermore, these equilibria are only {\it metastable} (but
long-lived) since the Boltzmann entropy has no global maximum. (ii)
Collisionless stellar systems (e.g. elliptical galaxies) can reach a
{\it metaequilibrium} state on a very short time scale $\sim t_D$ as a
result of a violent relaxation. This metaequilibrium state is a
stationary solution of the Vlasov equation. Lynden-Bell (1967) has
tried to predict this metaequilibrium state by resorting to a new type
of statistical mechanics. Unfortunately, his prediction is limited by
the problem of {incomplete relaxation} (e.g., Hjorth \& Madsen
1993). Therefore, H-functions must be introduced to account for the
structure of galaxies (Tremaine, H\'enon
\& Lynden-Bell 1987, Chavanis 2003a,c). It can be shown that the
maximization of a H-function
\begin{equation}
S=-\int C(f)d^{3}{\bf r}d^{3}{\bf v},
\label{mf6}
\end{equation}
at fixed $M$, $E$ and ${\bf L}$, where $C(f)$ is a convex
function (i.e. $C''(f)>0$) determines nonlinearly dynamically stable
stationary solutions of the Vlasov equation (Ipser \& Horwitz
1979). This is {\it similar} to a condition of microcanonical
stability in thermodynamics where $S$ plays the role of a generalized
entropy (Chavanis 2003a). Note that according to the Jeans theorem
(Binney \& Tremaine 1987), other solutions of the Vlasov equation can
be constructed which do not satisfy this criterion.  (iii) The
equilibrium properties of stars are obtained by coupling the
hydrostatic equations to a local equation of state. In simplest
models, it is assumed that the star is barotropic, i.e. $p=p(\rho)$.
We introduce the energy functional
\begin{eqnarray}
{\cal W}[\rho]=\int\rho\int_{0}^{\rho}{p(\rho')\over\rho'^{2}}d\rho'd^{3}{\bf r}+{1\over 2}\int \rho\Phi d^{3}{\bf r}\nonumber\\
+\int \rho {{\bf u}^{2}\over 2}d^{3}{\bf r}-{\bf\Omega}\int \rho\ {\bf r}\times {\bf u}\ d^{3}{\bf r},
\label{bs3}
\end{eqnarray}
where the first term is the internal energy, the second the
gravitational energy, the third the kinetic energy of the bulk motion
and the third the energy of rotation. It can be shown that the
minimization of ${\cal W}[\rho]$ at fixed mass $M$ and angular velocity
${\bf\Omega}$ determines nonlinearly dynamically stable stationary
solutions of the Euler-Jeans equations. This is {\it similar} to a
condition of canonical stability in thermodynamics where ${\cal W}$ plays the role of a generalized free energy (Chavanis
2003a). (iv) Finally, the interstellar medium can be considered as a
gas in thermal equilibrium with a radiation background (de Vega \&
Sanchez 2002). Therefore, for this system, the canonical description
makes sense. The stability of the gas against gravitational collapse
can be studied thermodynamically by minimizing the free energy
$F=E-TS$ associated with the Boltzmann entropy (\ref{mf6b}) or
dynamically by using the Jeans instability criterion (see Chavanis
2002e).  This is the traditional approach to understand star
formation.

In this paper, we shall consider a system of self-gravitating
fermions. They can be electrons in white dwarf stars, neutrons in
neutron stars or massive neutrinos in dark matter models. Their
statistical equilibrium state is obtained by maximizing the
Fermi-Dirac entropy
\begin{equation}
S=-\int \biggl\lbrace {f\over\eta_{0}}\ln{f\over\eta_{0}}+\biggl (1- {f\over\eta_{0}}\biggr )\ln \biggl (1-{f\over\eta_{0}}\biggr )\biggr\rbrace d^{3}{\bf r}d^{3}{\bf v},
\label{mf7}
\end{equation}
at fixed mass, energy and angular momentum. 
In this
formula  $\eta_{0}=(2s+1)m^{4}/(2\pi
\hbar)^{3}$ is the maximum allowable value of the distribution
function fixed by Pauli's exclusion principle ($s$ is the spin). For
the quantum gas, we shall consider either microcanonical or canonical
equilibria. Incidentally, the Fermi-Dirac entropy (\ref{mf7}) also
occurs in the statistical theory of violent relaxation for
collisionless stellar systems developed by Lynden-Bell (1967). It
corresponds to the two-levels approximation  $f_{0}\in
\lbrace 0,\eta_{0}\rbrace $ of his theory, where $f_{0}$ is the 
initial distribution function (Chavanis \& Sommeria 1998). In the non
degenerate limit $f\ll\eta_{0}$, the Fermi-Dirac distribution
(\ref{mf7}) returns the Boltzmann entropy (\ref{mf6b}) as a special
case. Therefore, our study can apply either to quantum particles,
stellar systems or to the interstellar medium. Of course, each system
corresponds to a different limit of the theory characterized by a
different degree of degeneracy and rotation.

Looking for critical points of entropy (\ref{mf7}) at
fixed mass, energy and angular momentum and introducing Lagrange
multipliers $\alpha$, $\beta$ and $-\beta{\bf\Omega}$ for each
constraint, we obtain the Fermi-Dirac distribution function
\begin{equation}
f={\eta_{0}\over 1+\lambda e^{\beta (\epsilon-{\bf \Omega}\cdot {\bf j})}},
\label{mf8}
\end{equation}
where $\epsilon={v^{2}\over 2}+\Phi$ and ${\bf j}={\bf r}{\times} {\bf
v}$ are the energy and the angular momentum of a particle by unit of
mass and $\beta= 1/T$ is the inverse temperature (we include $m$ or
$\eta_{0}$ in the definition of $T$). The same distribution
(\ref{mf8}) is obtained by extremizing the free energy $J=S-\beta
E+{\beta}{\bf\Omega}\cdot {\bf L}$ at fixed $\beta$, ${\bf\Omega}$ and
$M$. However, the second order variations of $S$ and $J$ (under
respective constraints) can have a different sign implying an
inequivalence of statistical ensembles (Chavanis 2003c). Introducing
the Jacobi energy $\epsilon_{J}\equiv
\epsilon-{\bf\Omega}\cdot {\bf j}$ and noting that
$\epsilon_{J}={w^{2}\over 2}+\Phi_{eff}$, where ${\bf w}={\bf
v}-{\bf\Omega}{\times} {\bf r}$ is the relative velocity and
$\Phi_{eff}=\Phi-{1\over 2}({\bf\Omega}{\times} {\bf r})^{2}$ is the
effective potential accounting for inertial forces, we can rewrite
Eq. (\ref{mf8}) in the form
\begin{equation}
f={\eta_{0}\over 1+\lambda e^{\beta ({w^{2}\over 2}+\Phi_{eff})}}.
\label{mf9}
\end{equation}
We note that the ``most probable'' form of rotation is a rigid rotation
${\bf\Omega}$. The gravitational potential $\Phi$ is defined within an
arbitrary gauge constant. We shall take the ordinary convention
$\Phi\rightarrow 0$ at infinity.

By integrating Eq. (\ref{mf9}) over the velocity, we find that the
density is given by
\begin{equation}
\rho={4\pi\sqrt{2}\eta_{0}\over \beta^{3/2}}I_{1/2}(\lambda e^{\beta\Phi_{eff}}),
\label{mf10}
\end{equation}
where $I_{n}$ is the Fermi integral
\begin{equation}
I_{n}(t)=\int_{0}^{+\infty}{x^{n}\over 1+t e^{x}}dx.
\label{mf11}
\end{equation}
The equilibrium configuration is then obtained by solving the Fermi-Poisson equation
\begin{equation}
\Delta\Phi={16\pi^{2}\sqrt{2}G\eta_{0}\over \beta^{3/2}}I_{1/2}(\lambda e^{\beta\Phi_{eff}}),
\label{mf12}
\end{equation}
and relating the Lagrange multipliers $\lambda$, $\beta$ and ${\bf
\Omega}$ to the constraints $M$, $E$ and ${\bf L}$. We stress that,
for long-range systems, the mean-field approximation is {\it exact} in
a suitable thermodynamic limit (see Sec. \ref{sec_tl}) so that our
``thermodynamical'' approach based on the maximization of
thermodynamical potentials is rigorous and simpler than the
``statistical'' approach based on functional integration (Laliena
1999, Votyakov {et al.} 2002). The relation between the two points of
view is discussed in Chavanis (2003a).

\subsection{The box model}
\label{sec_box}

It is easy to see that Eq. (\ref{mf12}) has no physical
solution. Indeed, the density decreases as $r^{-2}$ at large distances
so that the total mass of the configuration is infinite in
contradiction with our starting hypothesis. This means that there is
no maximum entropy state in an unbounded domain. A self-gravitating
system at non-zero temperature has the tendency to ``evaporate''.
Therefore, the statistical mechanics of self-gravitating systems
(classical or quantum) is essentially an out-of-equilibrium
problem. However, evaporation is a slow process and, in practice, the
system is found in a {\it quasi-equilibrium} state with a
``core-halo'' structure. The core of the system is approximately
isothermal while the halo departs from isothermality. Therefore, the
distribution function (\ref{mf8}) must be modified at high
energies. Some modified isothermal distributions have been introduced
in the case of globular clusters (Michie 1963, King 1966) subject to
the tides of a nearby object (typically a spiral galaxy) and in the
case of elliptical galaxies undergoing {\it incomplete} violent relaxation
(Hjorth \& Madsen 1993).

Another astrophysically plausible distribution
function which takes into account Pauli's exclusion principle and
tidal effects is given by (Chavanis 1998)
\begin{equation}
f=\eta_{0}{e^{-\beta\epsilon}-e^{-\beta\epsilon_{m}}\over \lambda+e^{-\beta\epsilon}},
\label{box1}
\end{equation}
for $\epsilon\le \epsilon_{m}$ and $f=0$ for $\epsilon\ge
\epsilon_{m}$ (where $\epsilon_{m}$ is the escape energy). This distribution
function can be derived from a Fokker-Planck equation appropriate to
self-gravitating fermions. It can describe dark matter halos limited
in extension by tidal effects (and, possibly, certain collisionless stellar
systems). In the classical limit, it reduces to the Michie-King model.
The distribution function (\ref{box1}) will be studied specifically in
another paper. For the present, we shall keep the full Fermi-Dirac
distribution function (\ref{mf9}) and confine the system within a
spherical box of radius $R$ so as to avoid the infinite mass problem
while preserving the rotational symmetry of the system. Typically, the
size of the box delimitates the physical extent of the system under
consideration. Putting the system in a box is a simple and crude way
of handling the infinite mass problem. This is sufficient, in a first
step, if we are just interested in describing phase transitions
between gaseous and ``core-halo'' states. If, on the other hand, we
wish to make more realistic models of self-gravitating systems, the
spatial confinement must be given further consideration.

\subsection{Asymptotic limits $T=0$ and $T\rightarrow +\infty$}
\label{sec_as}

It will be convenient in the following to work with dimensionless
variables defined by
\begin{eqnarray}
{\bf x}={{\bf r}\over R},\quad  n={\rho\over (M/R^{3})}, \quad \phi={\Phi\over (GM/R)},\nonumber\\
\Lambda=-{ER\over GM^{2}}, \quad \eta={\beta GM\over R},\quad \Omega'=\Omega \biggl ({R^{3}\over GM}\biggr )^{1/2},\nonumber\\
L'={L\over\sqrt{GM^{3}R}},\quad \mu=\eta_{0}\sqrt{512\pi^{4}G^{3}MR^{3}}.
\label{as1}
\end{eqnarray}
This is equivalent to setting $M=R=G=1$ in the dimensional equations
and we shall adopt this convention in the following. Then, the
degeneracy parameter is given by $\mu=16\sqrt{2}\pi^{2}\eta_{0}$. We
also define $\psi=\beta\Phi$ and $\omega=\sqrt{\beta}\Omega$. Denoting
by $\psi_{0}$ the value of $\psi$ at $r=0$ and introducing the
uniformizing variable $k=\lambda e^{\psi_{0}}$, the Fermi-Poisson
equation (\ref{mf12}) becomes
\begin{equation}
\Delta\psi=\mu\sqrt{T}I_{1/2}\bigl (k e^{\psi-\psi_{0}-{1\over 2}\omega^{2}s^{2}}\bigr ),
\label{as2}
\end{equation}
where $s=r\sin\theta$ denotes the distance to the axis of
rotation. Equation (\ref{as2}) must be solved with the condition
$\psi\rightarrow 0$ at infinity. Before solving this equation in the
general case, we first discuss important asymptotic limits.

For $k\rightarrow +\infty$, we can use the limiting form of the Fermi integral
\begin{equation}
I_{n}(t)\sim {1\over t}\Gamma(n+1),\qquad (t\rightarrow +\infty),
\label{as3}
\end{equation}
and we obtain the Boltzmann-Poisson equation
\begin{equation}
\Delta\psi={\mu\sqrt{\pi T}\over 2k} e^{-\psi+\psi_{0}+{1\over 2}\omega^{2}s^{2}},
\label{as4}
\end{equation}
for a rotating isothermal gas. This corresponds to the high
temperature limit of the Fermi distribution. Indeed, for $T\gg 1$ the distribution function (\ref{mf9}) is Maxwellian
\begin{equation}
f={\eta_{0}\over \lambda}e^{-\beta({w^{2}\over 2}+\Phi_{eff})}.
\label{as5}
\end{equation}
Alternatively, for $k\rightarrow 0$,
we can use the limiting form of the Fermi integral
\begin{equation}
I_{n}(t)\sim {(-\ln t)^{n+1}\over n+1},\qquad (t\rightarrow 0),
\label{as6}
\end{equation}
and obtain the Lane-Emden equation
\begin{equation}
\Delta\psi={2\over 3}\mu\sqrt{T}\bigl (-\ln k-\psi+\psi_{0}+{1\over 2}\omega^{2}s^{2}\bigr )^{3/2},
\label{as7}
\end{equation}
for a rotating polytrope of index $n=3/2$. This corresponds to the low
temperature limit of the Fermi distribution in which the structure is
completely degenerate (white dwarf). Indeed, for $T=0$, the
distribution function (\ref{mf9}) is a step function
\begin{equation}
f=\eta_{0}H(\epsilon_{J}-\epsilon_{F}),
\label{as8}
\end{equation}
where $\epsilon_{J}$ is the Jacobi energy, $\epsilon_{F}=-(1/\beta)\ln\lambda$ is the Fermi energy and $H$ is the Heaviside function. Therefore, the Fermi-Dirac distribution function connects continuously isothermal and polytropic
distributions for high and low temperatures.

We recall that for the self-gravitating Fermi gas at $T=0$ (white
dwarf), the density vanishes at a finite radius $R_*$. The mass-radius
relation can be written (Chandrasekhar 1942)
\begin{equation}
MR_{*}^{3}={\chi\over \eta_{0}^{2}G^{3}},
\label{req1}
\end{equation}
with $\chi\simeq 5.9723\ 10^{-3}$. The condition $R_{*}\le R$ is equivalent to
\begin{equation}
\mu\ge \mu_{*}\equiv \sqrt{512\pi^{4}\chi}\simeq 17.259.
\label{req2}
\end{equation}
More generally, we have
\begin{equation}
\mu=\mu_{*}\biggl ({R\over R_{*}}\biggr )^{3/2}.
\label{req3}
\end{equation}
Therefore, the degeneracy parameter $\mu$ is related to the ratio of
the system size $R$ to the radius $R_{*}$ of a white dwarf star with
mass $M$. Accordingly, a small value of $\mu$ corresponds to a large
``effective'' cut-off (played by Pauli's exclusion principle) or,
equivalently, to a small system size. Alternatively, a large value of
$\mu$ corresponds to a small ``effective'' cut-off or a large system
size.

\subsection{The thermodynamic limit}
\label{sec_tl}

Before going further, we shall discuss the thermodynamic limit of
self-gravitating systems. In order to precisely define the
thermodynamic limit, we must introduce a small-scale regularization
otherwise the density of states $g(E)$ diverges (Padmanabhan
1990). The most physical regularization is to consider quantum
mechanics effects. Indeed, compact objects such as white dwarf stars
and neutron stars owe their stability to quantum mechanics. The
thermodynamic limit corresponds to $N\rightarrow +\infty$ such that
$\Lambda=-ER/GN^{2}m^{2}$, $\eta=\beta GNm/R$,
$\Omega'=\Omega(R^{3}/GNm)^{1/2}$, $L'=L/\sqrt{GN^{3}m^{3}R}$ and
$\mu=(gm^{4}/h^{3})\sqrt{512\pi^{4}G^{3}NmR^{3}}$ are kept fixed.  It
is natural to fix $m$, $h$ and $G$. Then, the strict thermodynamic limit is
$N\rightarrow +\infty$ with
\begin{eqnarray}
R\sim N^{-1/3}, \quad E\sim N^{7/3},\quad \beta \sim N^{-4/3},\nonumber\\
\Omega\sim N, \quad L\sim N^{4/3}. \qquad\qquad\qquad\qquad
\label{gtq1}
\end{eqnarray}
This scaling was indicated in a previous paper (Chavanis 2002b).  When
this limit is taken, $S\sim N$, $J\sim N$ and the mean-field
approximation is exact (note that the usual free energy
$F=E-TS=-J/\beta$ scales as $N^{7/3}$). The scaling $MR^{3}\sim 1$
corresponds to the ``white dwarf'' mass-radius relation
(\ref{req1}). The fact that $R\rightarrow 0$ as $N\rightarrow +\infty$
accounts for the natural tendency of self-gravitating systems to
collapse (note that when $M$ reaches Chandrasekhar's limiting mass,
Newtonian mechanics loses its relevance). We shall call (\ref{gtq1})
the quantum, or white dwarf, thermodynamical limit (QTL). It is
particularly relevant for compact objects forming the condensed phase.

If we now consider the classical limit $h=0$, we see that $\mu$
becomes infinite so that it disappears from the problem. Then, the
classical thermodynamic limit (CTL) is $N\rightarrow +\infty$ with
fixed $\Lambda=-ER/GN^{2}m^{2}$, $\eta=\beta GNm/R$,
$\Omega'=\Omega(R^{3}/GNm)^{1/2}$, $L'=L/\sqrt{GN^{3}m^{3}R}$. We note
that there is a freedom since we have to fix a parameter in order to
determine the others. If we assume $\beta\sim 1$, the classical
thermodynamic limit is $N\rightarrow +\infty$ with
\begin{eqnarray}
R\sim N, \quad E\sim N,\quad \beta \sim 1,\qquad\qquad\qquad\nonumber\\
\Omega\sim N^{-1}, \quad L\sim N^{2}. \qquad\qquad\qquad\qquad
\label{gtq1Class}
\end{eqnarray}
This has been called the ``dilute limit'' by de Vega \& Sanchez
(2002). Again, $S\sim N$, $J\sim N$ (and now $F\sim N$) so that the
mean-field approximation is exact in that limit. We recall that for
$h=0$, the density of state diverges due to the existence of collapsed
configurations. However, for sufficiently large $T$ and $E/N$, we know
that there exists long-lived {\it metastable} gaseous states (local
entropy maxima).  For $N\gg 1$, these metastable states have a very
long lifetime, increasing as ${\rm exp}(N)$, so they constitute the most
relevant astrophysical structures (Katz \& Okamoto 2000, Chavanis
\& Ispolatov 2002, Chavanis 2003a). Globular clusters are in such metastable
states. These gaseous states are independent on the small-scale
cut-off so they are described adequately by the dilute limit
(\ref{gtq1Class}). The consideration of these metastable states may
solve the problems raised by Laliena (2003). Collapsed configurations
(binaries in MCE and Dirac peaks in CE), which correspond to global
maxima of entropy or free energy (i.e., true equilibrium states) must
be discarded ``by hands'' because they are reached for unphysically
large times $t\rightarrow +\infty$. They can be reached for accessible
times only if the temperature and the energy are sufficiently small
($T<T_c$ or $E<E_c$) because, in that case, the gaseous phase ceases
to exist and the system undergoes gravitational collapse. Clearly, the
end-state of the collapse depends on the nature of the small-scale
cut-off (e.g., quantum degeneracy) and we must consider the
thermodynamical limit (\ref{gtq1}) if we want to describe these
condensed states.

If instead of fermions we consider a hard sphere gas, the filling
factor is $\mu=R/aN^{1/3}$ where $a$ is the hard sphere radius
(see Chavanis 2002). Then, fixing $a$, $m$ and $G$, the
strict thermodynamic limit is $N\rightarrow +\infty$ with
\begin{eqnarray}
R\sim N^{1/3}, \quad E\sim N^{5/3},\quad  \beta \sim N^{-2/3},\nonumber\\
 \Omega\sim 1, \quad L\sim N^{5/3}.\qquad\qquad\qquad\qquad
 \label{gtq2}
\end{eqnarray}
Finally, if we consider a soften potential with
soften parameter $r_{0}$, the cut-off parameter is $\mu=R/r_{0}$
(see Chavanis \& Ispolatov 2002).  Then, fixing $r_{0}$, $m$ and $G$, the
strict thermodynamic limit is $N\rightarrow +\infty$ with
\begin{eqnarray}
R\sim 1,\quad  E\sim N^{2},\quad  \beta \sim N^{-1}, \nonumber\\
\Omega\sim N^{1/2}\quad L\sim N^{3/2}.\qquad\qquad
\label{gtq3}
\end{eqnarray}
The dilute limit (\ref{gtq1Class}) is of course unchanged and it
describes gaseous metastable states independent on the small-scale
cut-off. Note that other combinations keeping $\Lambda$, $\eta$,
$\Omega'$ and $L'$ fixed can be considered (e.g., $G\sim 1/N$ for
fixed $m$, $R$, $\beta$, $E/N$, $\Omega$ and $L/N$ or $m\sim 1/N$ for
fixed $G$, $R$, $\beta$, $E$, $\Omega$ and $L$). Of course, the limits
considered above are only formal. In reality, one does not have to
make $N$, $R$,... go to $+\infty$ or $0$. The precise values taken by
these parameters are fixed by the physics of the system under
consideration (molecular clouds, globular clusters, white dwarfs,
neutron stars, dark matter halos,...). Note that even for small values
of $N$, the mean-field approximation is accurate so that the
thermodynamic limit makes sense even if we are not in the strict,
mathematical, conditions of its application.

\subsection{Integral constraints}
\label{sec_ic}

According to the classical work of James (1964), rotating polytropes
with index $n>0.808$ do not bifurcate to
non-axisymmetric structures when rotation is increased. Instead, they
develop a cusp at the equator. Since the Fermi gas at $T=0$ is
equivalent to a polytrope of index $n=3/2$, it is not expected to
bifurcate to non-axisymmetric structures (in contrast to homogeneous
bodies). We expect this property to persist at non-zero temperature
since, at $T\neq 0$, we have essentially a polytrope $3/2$ surrounded
by a dilute halo. We do not exclude the possibility of {\it
discontinuous} bifurcations to non-axisymmetric structures or
rings. However, in this paper, we shall restrict ourselves to single-cluster
axisymmetric solutions of Eq. (\ref{as2}). 

In that case, the angular momentum can be written ${\bf
L}=I{\bf\Omega}$ where $I$ is the moment of inertia
\begin{equation}
I=2\pi \int_{-1}^{+1}\int_{0}^{1}\rho r^{4}(1-\mu^{2})dr d\mu,
\label{ic1}
\end{equation}
where $\mu=\cos\theta$. The energy can be deduced from the Virial theorem (see, e.g., Chavanis 2002c)
\begin{equation}
E=-K+\oint p {\bf r}\cdot d{\bf S}.
\label{ic2}
\end{equation}
The kinetic energy $K$ incorporates a rotational contribution
\begin{equation}
K_{rot}={1\over 2}{\bf L}\cdot {\bf\Omega}={1\over 2}I\Omega^{2},
\label{ic3}
\end{equation}
and a thermal contribution
\begin{equation}
K_{th}={3\over 2}\int p d^{3}{\bf r},
\label{ic4}
\end{equation}
where $p={1\over 3}\int fw^{2}d^{3}{\bf w}$ is the local
pressure. Using Eq. (\ref{mf9}), it can be written
\begin{equation}
p={8\pi\sqrt{2}\eta_{0}\over 3 \beta^{5/2}}I_{3/2}(\lambda
e^{\beta\Phi_{eff}}),
\label{ic5a}
\end{equation}
or, in dimensionless form,
\begin{equation}
p={\mu\over 6\pi}T^{5/2}I_{3/2}(ke^{\psi-\psi_{0}-{1\over 2}\omega^{2}s^{2}}).
\label{ic5}
\end{equation}
Equations (\ref{mf10}) and (\ref{ic5a}) determine the equation of state
of the Fermi gas at non-zero temperature (in a parametric form). Since
$p=p(\rho)$, this gas is {\it barotropic}.

\subsection{A short description of the numerical procedure}
\label{sec_np}

Using the variables introduced previously, the Gauss theorem can be written as
\begin{equation}
\oint \nabla\psi\cdot d{\bf S}={4\pi\over T},
\label{np}
\end{equation}
where the surface integral is computed on the spherical box. Now, using
the decomposition of the gravitational potential in normalized
spherical harmonics, i.e.,

\begin{equation}
\psi=\sum_{l,m}\psi_{lm}(r)Y_{lm}(\theta,\phi),
\label{np1}
\end{equation}
it turns out that
\begin{equation}
T={4\pi\over \psi'_{00}(1)}.
\label{np2}
\end{equation}
Therefore, Eq. \eq{as2} depends only on three parameters $\mu$, $\Omega$
and $k$. The degeneracy parameter $\mu$ and the angular velocity
$\Omega$ specify a series of equilibria parameterized by the
uniformizing variable $k$. Given these three parameters, the
Fermi-Poisson equation \eq{as2} can be solved by an iterative procedure. For
that purpose, the radial components of the potential, i.e.
$\psi_{lm}(r)$, are discretized on the Gauss-Lobatto grid which ensures
spectral convergence. Typically, the resolution used for the following
numerical results was 100 radial grid points and a decomposition on 16
spherical harmonics.

The temperature $T(\mu,\Omega,k)$, the energy $E(\mu,\Omega,k)$ and the
angular momentum $L(\mu,\Omega,k)$ are determined by Eqs.
\eq{np2}, \eq{ic2} and \eq{ic1}. It is then possible to compute the
caloric curves $E(T)$ for given $\mu$ and $\Omega$ (or $L$) by
eliminating $k$ from the previous relations.

\section{Caloric curves for low values of the degeneracy 
parameter (high cut-off/small system)}
\label{sec_fe}

We first describe the figures of equilibrium of the rotating
self-gravitating Fermi gas for a degeneracy parameter $\mu=100$. This
value is smaller than the microcanonical critical point
$\mu_{MTP}=2670$ above which a first order microcanonical phase
transition appears (Chavanis 2002b). Hence, for $\mu=100$ (small
system), the influence of the small-scale cut-off (played here by the
exclusion principle) is strong in average and stabilizes the system
with respect to the gravothermal catastrophe (Lynden-Bell \& Wood
1968).

\subsection{Non-rotating configurations}
\label{sec_nr}

The caloric curve of non-rotating self-gravitating fermions has been
discussed by Chavanis (2002b) in both canonical and microcanonical
ensembles and for arbitrary values of the degeneracy parameter $\mu$
(or system size $R$). This completes the work of Hertel \& Thirring
(1971) who worked in canonical ensemble and considered small system
sizes. For non-rotating systems, the results obtained with fermions
are similar to those obtained with other small-scale regularizations,
the small-scale cut-off playing the same role as the inverse of the
degeneracy parameter (Stahl et al. 1994, Follana \& Laliena 2000,
Chavanis \& Ispolatov 2002).

\begin{figure}[htbp]
\centerline{
\includegraphics[width=8cm,angle=0]{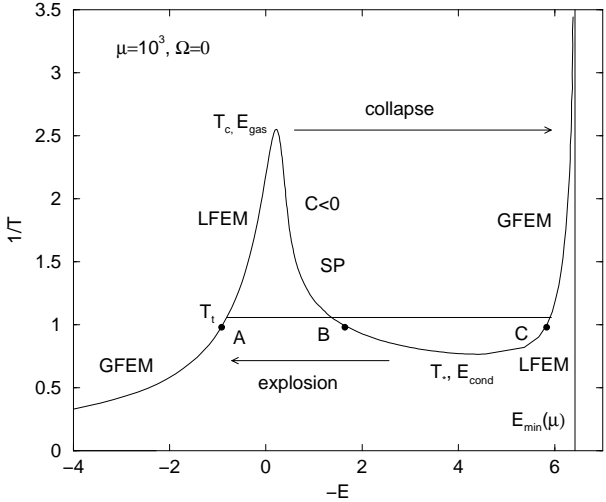}
}
\caption[]{The caloric curve for a non-rotating system with $\mu=1000<\mu_{MTP}=2670$. In the microcanonical ensemble (fixed $E$), the system slowly evolves
from the gaseous phase to the condensed phase, passing by a mixed
phase, as energy decreases. In the canonical ensemble (fixed $T$), the
system undergoes a sharp phase transition at a critical temperature
(see text).  Below $T_c$, the gaseous states undergo a gravitational
collapse (Jeans instability); above $T_*$, the condensed states
undergo an explosion. Between, $T_{t}$ and $T_{c}$ (resp. $T_{*}$) the
gaseous (resp. condensed) states are {\it metastable}. By varying the
temperature of the system between $T_*$ and $T_c$ (spinodal points),
we can generate an hysteretic cycle between the gaseous and the
condensed phase. }
\label{calor0}
\end{figure}

\begin{figure*}
\centerline{
\includegraphics[width=5cm,angle=0]{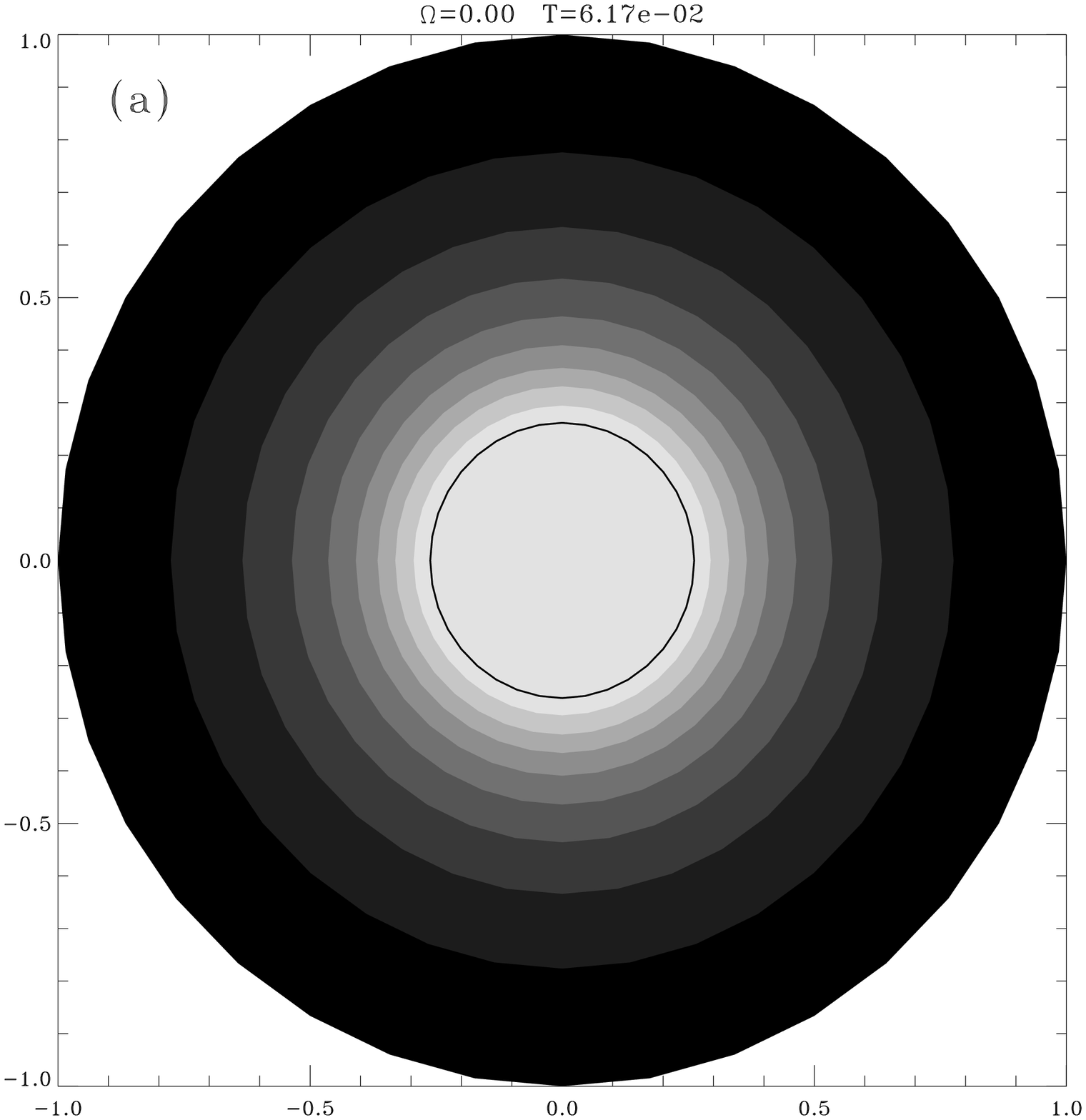}
\includegraphics[width=5cm,angle=0]{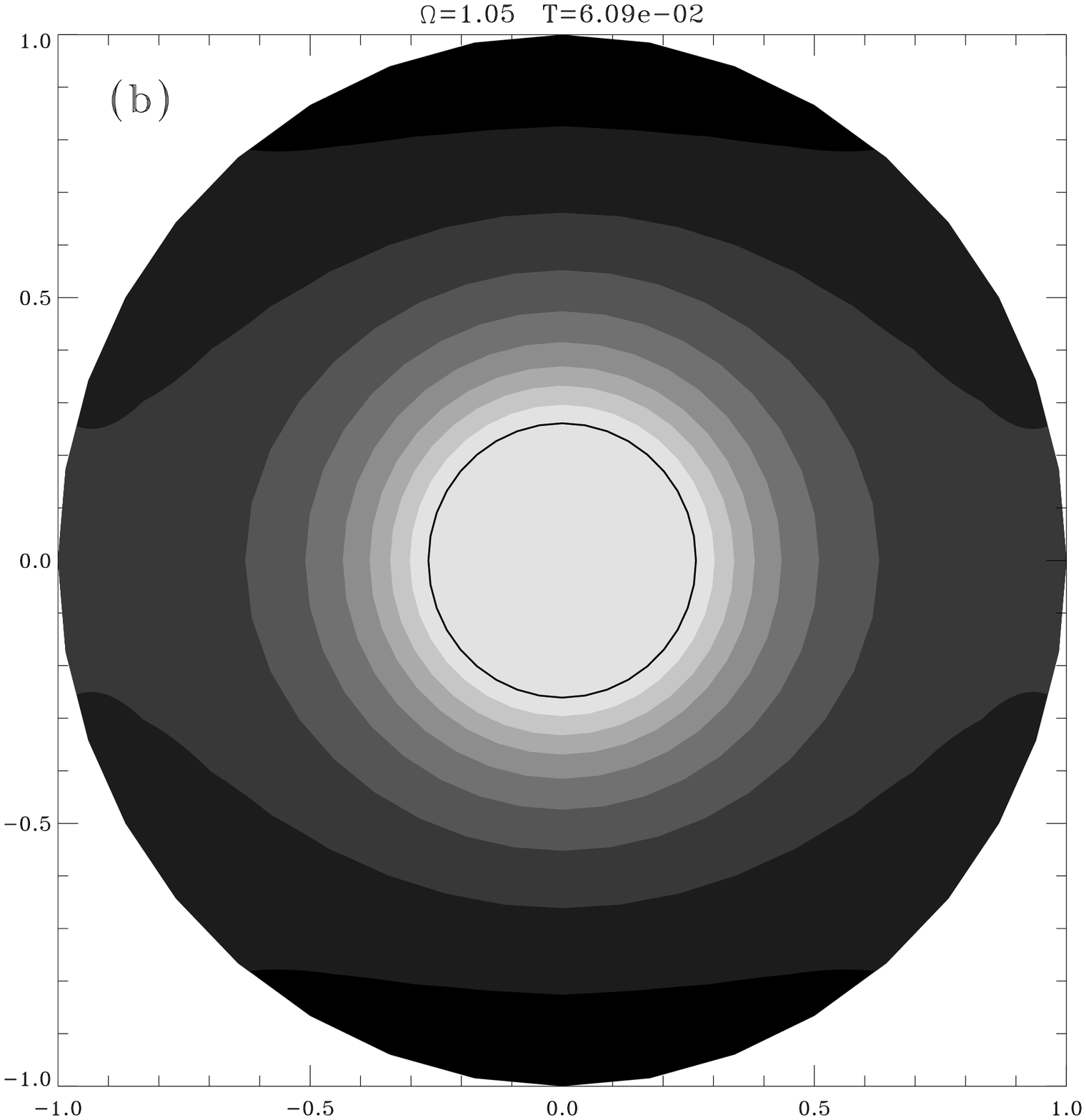}
\includegraphics[width=5cm,angle=0]{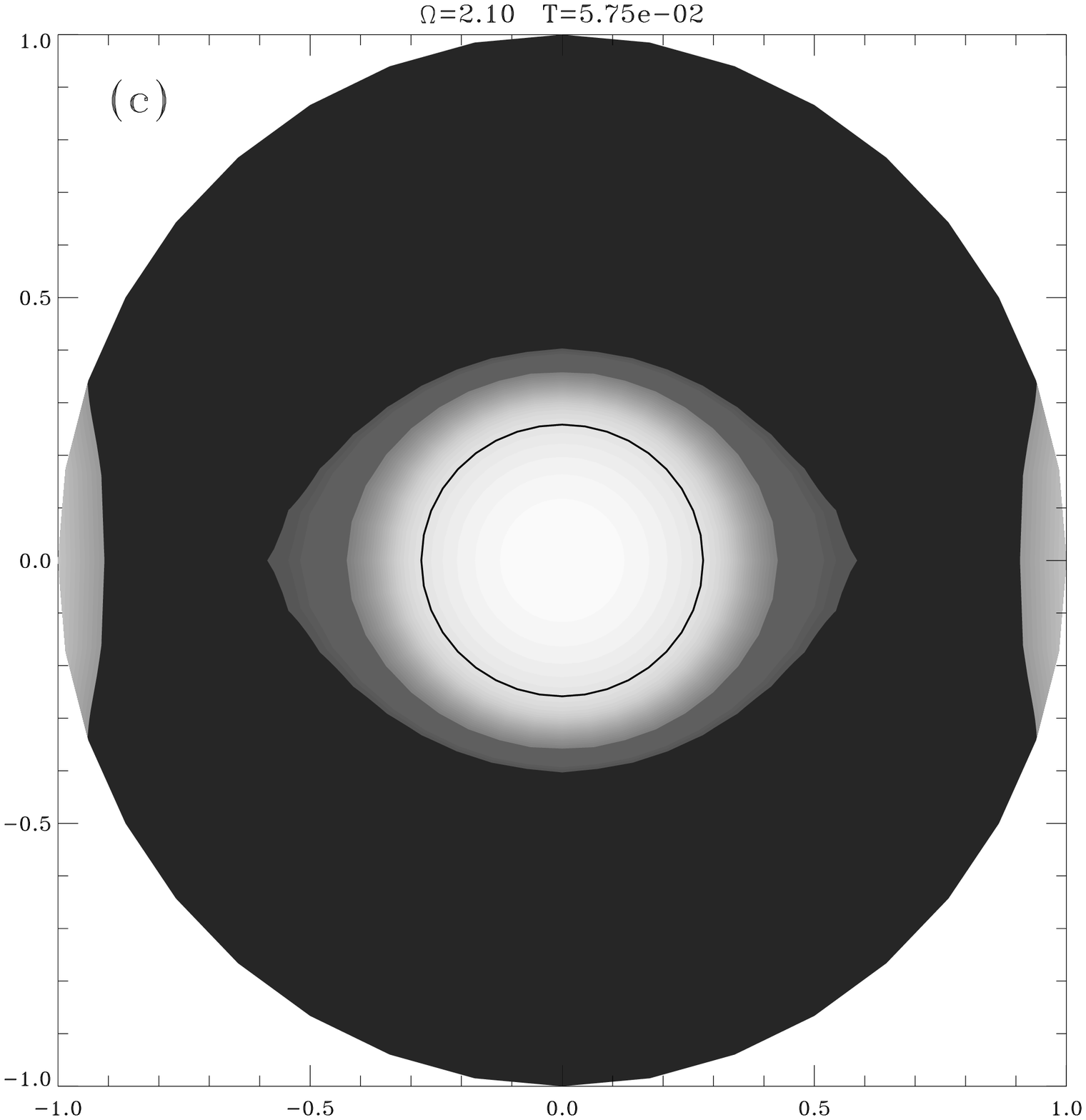}
}
\centerline{
\includegraphics[width=5cm,angle=0]{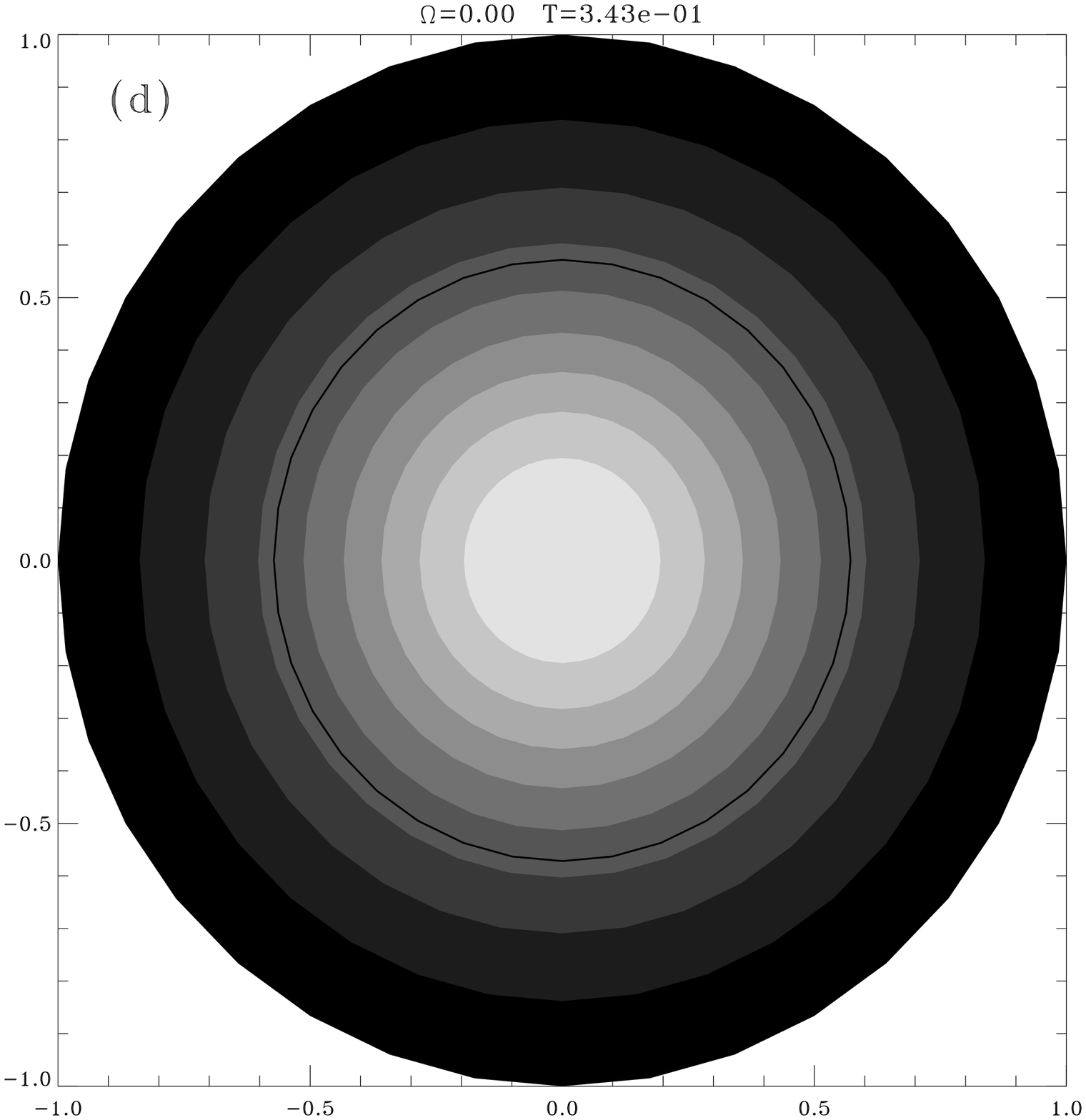}
\includegraphics[width=5cm,angle=0]{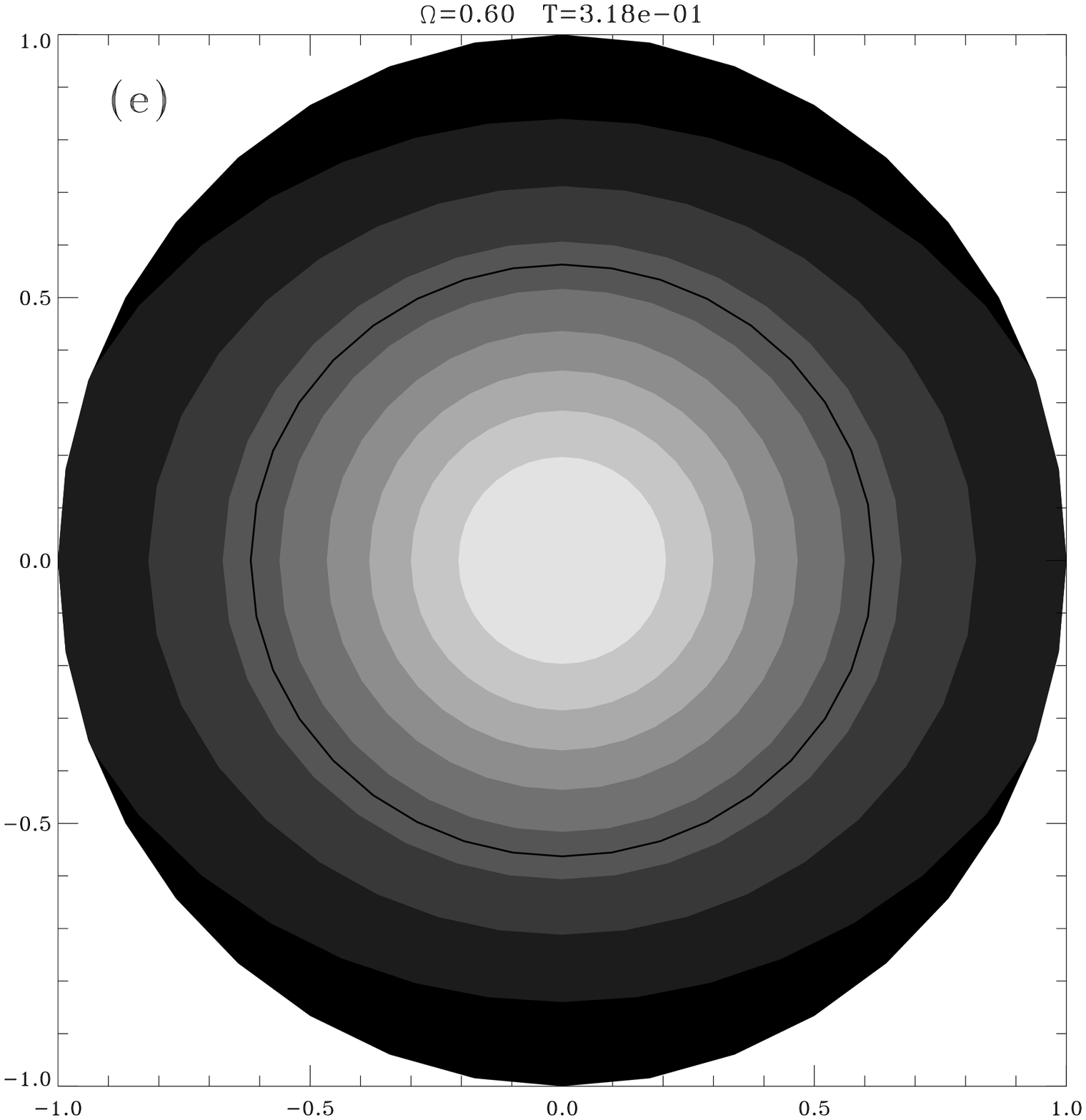}
\includegraphics[width=5cm,angle=0]{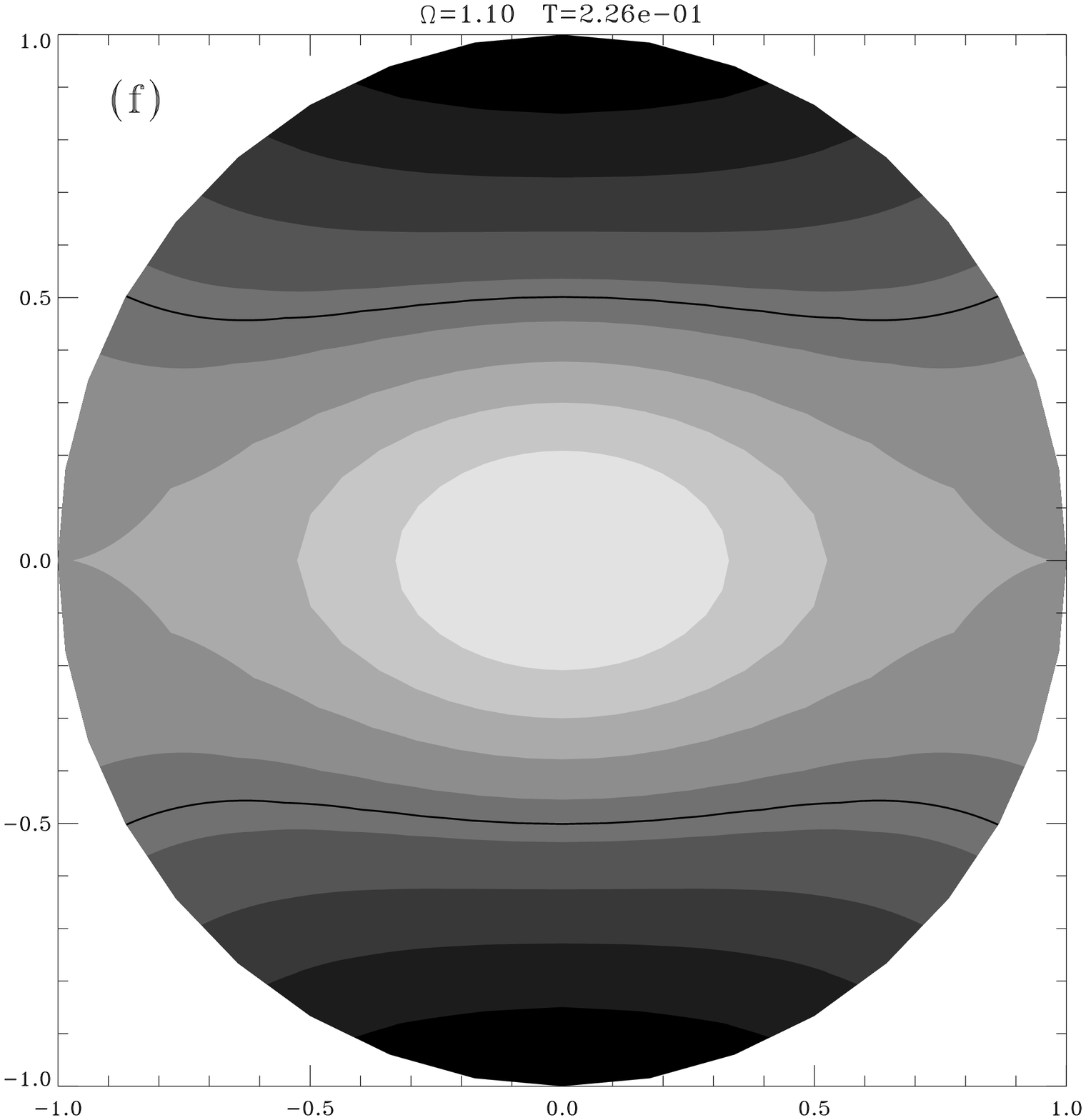}
}
\centerline{
\includegraphics[width=5cm,angle=0]{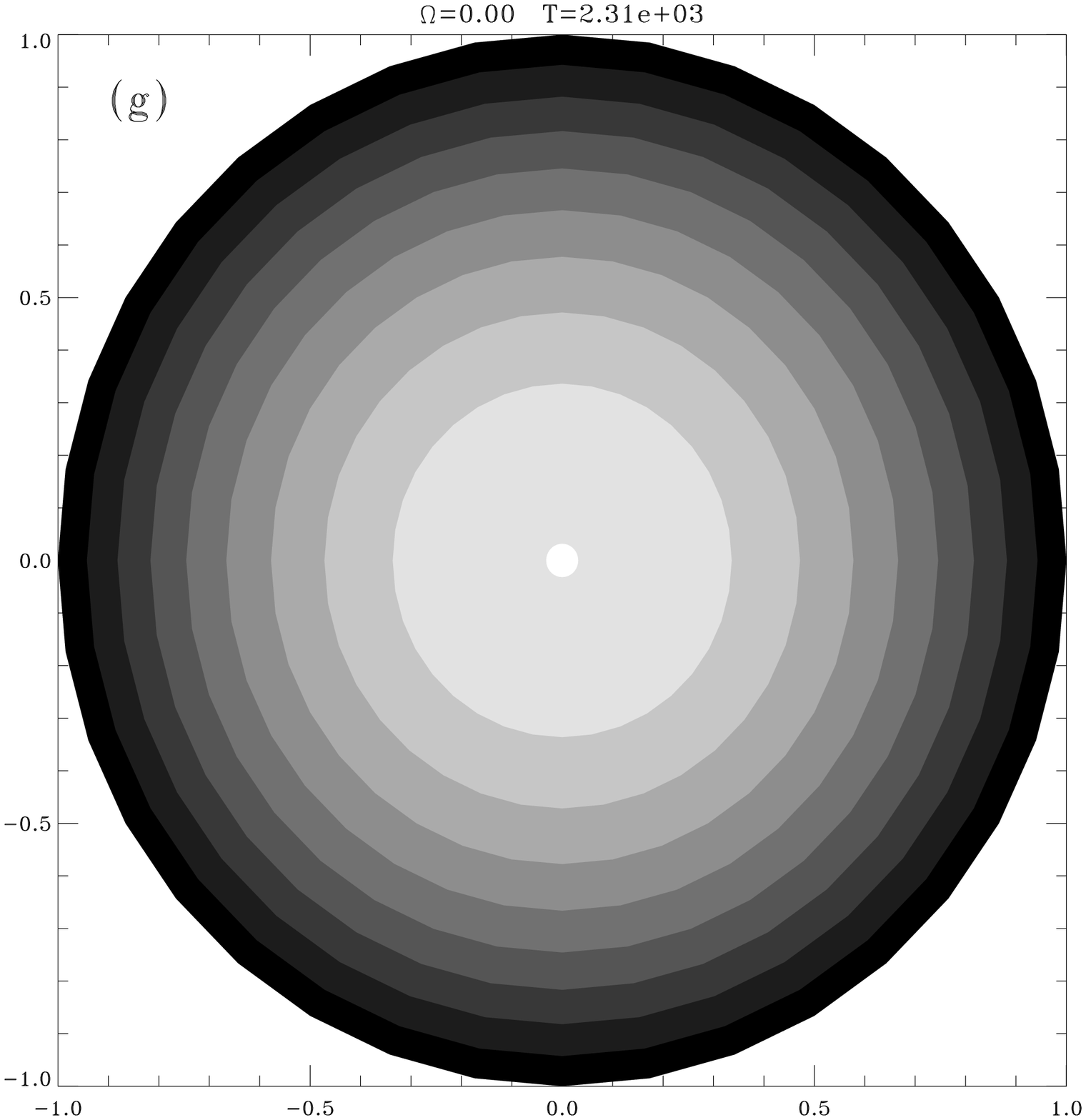}
\includegraphics[width=5cm,angle=0]{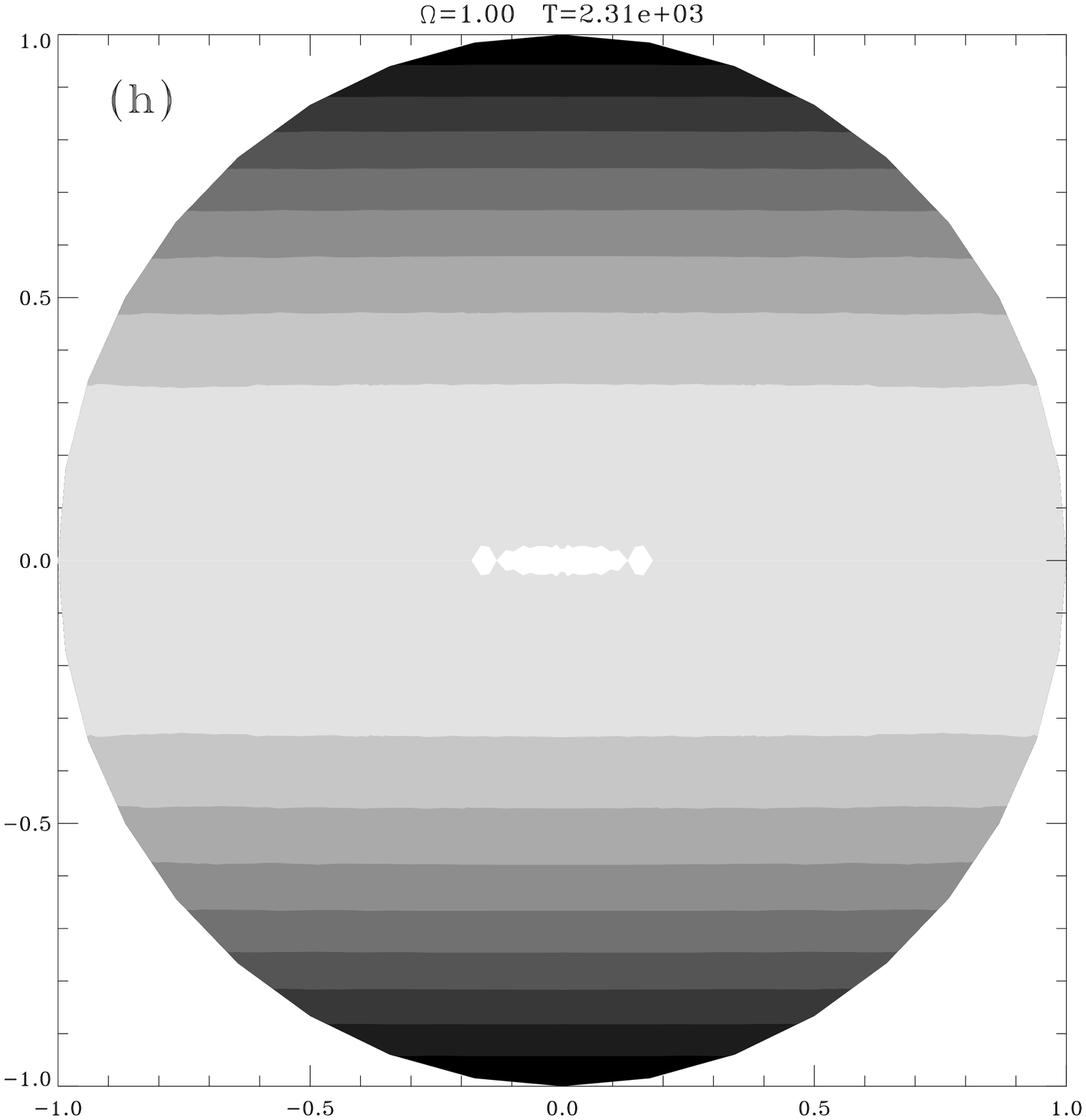}
\includegraphics[width=5cm,angle=0]{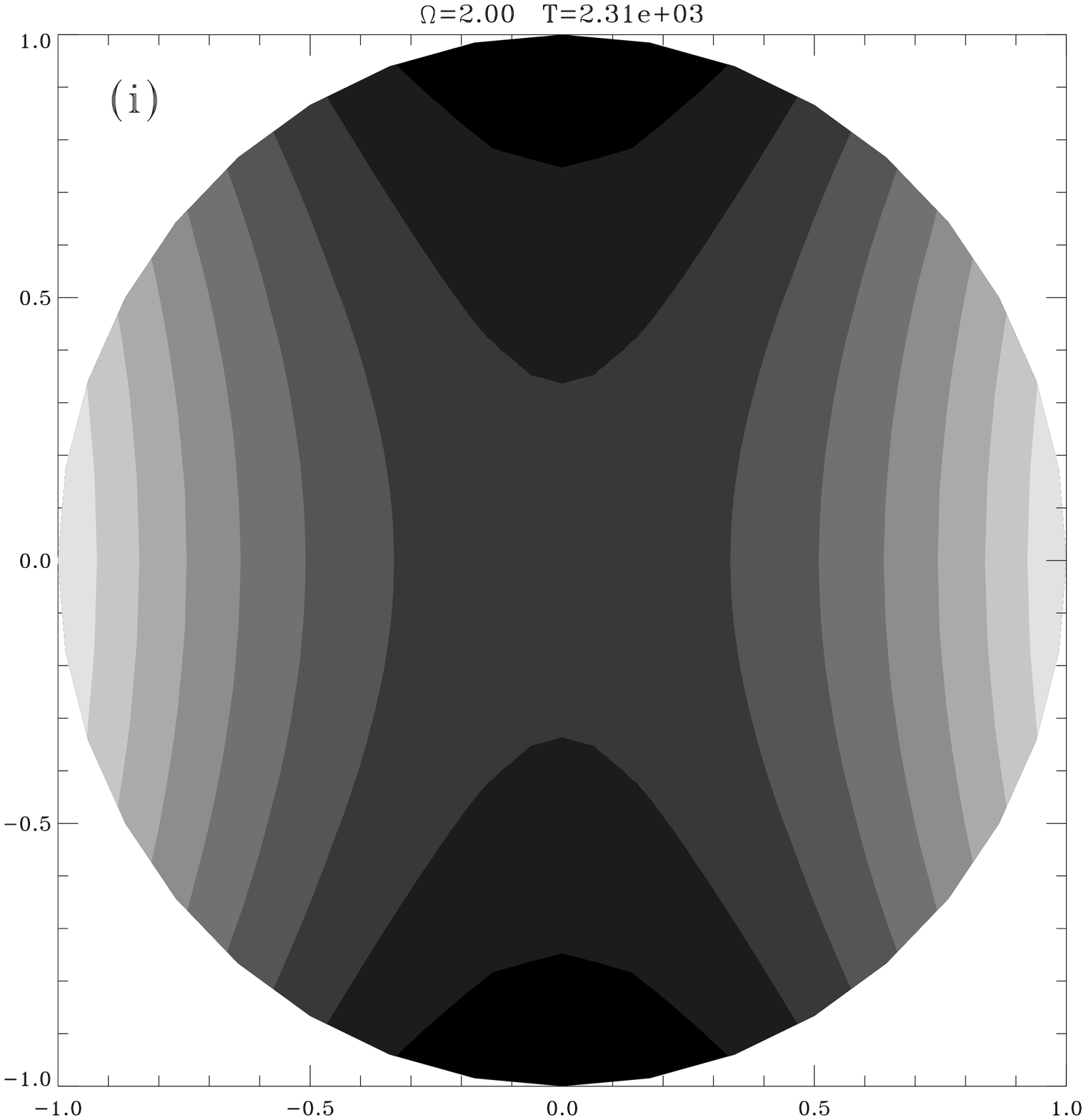}
}
\caption[]{Plot of the density (logarithmic scale) in a meridional
section of the box. The solid line marks the level where density is
0.05, the central density being unity. The first row shows
configurations at very low temperature/energy (condensed phase). The
second row shows configurations in the region of negative specific
heats (mixed phase). Such structures are forbidden (unstable) in the canonical
ensemble. The third row shows configurations at high
temperature/energy (gaseous phase). Alternatively, the first column
corresponds to non-rotating systems, the second column to slowly
rotating systems and the third column to systems rotating close to the
Keplerian limit.}
\label{density}
\end{figure*}

We first discuss the microcanonical ensemble corresponding to an
isolated system with fixed energy. For $\mu<\mu_{MTP}=2670$, there
exists only {one} extremum of entropy for each energy and it is a {\it
global} entropy maximum (GEM). The caloric curve is plotted in
Fig.~\ref{calor0} (in this figure $\mu=1000$). For high energies, the
density is almost homogeneous (Fig.~\ref{density}g). This forms the
``gaseous'' phase. For energies close to the minimum accessible energy
$E_{min}=-6.42\ 10^{-2}\mu^{2/3}$ (ground state), the temperature goes
to zero and the system is completely degenerate (Fig.~\ref{density}a).
It has the same structure as a classical white dwarf star, i.e. a
polytrope of index $n=3/2$. This forms the ``condensed'' phase. For
intermediate energies ($E_{cond}<E<E_{gas}$), corresponding to the
region of negative specific heats $C=dE/dT<0$, the system has a
``core-halo'' structure with a degenerate nucleus (fermion ball) and a
dilute isothermal atmosphere (Fig.~\ref{density}d) . This is like a
solid condensate embedded in a vapor in usual phase transitions. This
``mixed phase'' is intermediate between the pure condensed phase and
the pure gaseous phase.

We now describe the canonical ensemble in which the temperature $T$ is
fixed instead of the energy $E$. Thus, we need to find maxima of free
energy $J=S-\beta E$ at fixed $\beta$ and $M$. We recall that, for an
{\it isolated} self-gravitating system, the validity of using the
canonical ensemble to describe the statistics of a sub-system cannot
be established from the microcanonical ensemble since energy is not
additive (Padmanabhan 1990). Therefore, the notion of thermostat has
to be defined carefully. The canonical ensemble can make sense for
self-gravitating systems subject to additional short-range
interactions. They can constitute a thermostat imposing a
temperature. In that case, the usual canonical ensemble interpretation
holds. For example, a self-gravitating system in contact with a
radiation background (e.g., a molecular cloud) is usually treated in
the canonical ensemble. This is the traditional way of studying
stellar formation in relation with Jeans instability.  Note
that there exists a precise theoretical model where the canonical
ensemble is {rigorously} justified. This is the self-gravitating
Brownian gas model (Chavanis et al. 2002, Sire \& Chavanis 2002,
2003). In this model, the temperature has a precise physical
interpretation since it is related to the strength of the stochastic
force (``kicks'') acting on the particles.

In the canonical ensemble, the region of negative specific heat is
unstable as it corresponds to saddle points (SP) of free energy (Katz 1978,
Chavanis \& Sommeria 1998). It is replaced by a canonical first order
phase transition connecting the gaseous phase for $T>T_{t}(\mu)$ to
the condensed phase for $T<T_{t}(\mu)$.  Therefore, the mixed
``core-halo'' states with $C<0$ are thermodynamically forbidden in the
canonical ensemble.  The gaseous states with temperature
$T_{c}<T<T_{t}$ are metastable (local maxima of free energy LFEM) but they
are long-lived. They can play an important role reminiscent of a
supercooled state of the Van der Waals gas. At $T<T_{c}$ they undergo
a gravitational collapse (zeroth order phase transition) coinciding
with Jeans instability criterion (Chavanis 2002e). For quantum
particles, the outcome of this collapse is the formation of a
``fermion ball'' containing almost {\it all} the mass. In the classical limit
$\mu\rightarrow +\infty$, the isothermal collapse leads to a Dirac
peak (see Sec. \ref{sec_gc}). The condensed states with temperature
$T_{t}<T<T_{*}$ are also metastable and long-lived (if they are
initially prepared in such states). They will undergo an
``explosion'', reversed to the collapse, if they are heated above
$T_*$. This explosion transforms the dense core into a relatively
uniform mass distribution. By varying the temperature between $T_*$
and $T_c$ we can generate an {\it hysteretic cycle} in the canonical
ensemble. This hysteretic cycle was described qualitatively in
Chavanis (2003b). Recently, it has been followed numerically by using
a model of self-gravitating Brownian fermions (Chavanis et al. 2003).
The turning points of temperature where the metastable states terminate
are called (canonical) spinodal points in the langage of phase
transition.  The first order canonical phase transition disapears for
$\mu<\mu_{CTP}=83$ (canonical critical point), i.e. for very high
cut-off values. This is simply because the small-scale ``repulsion''
prevails over gravity. More details on phase transitions in
self-gravitating systems can be found in Chavanis (2002b).

\subsection{Slowly rotating configurations}
\label{sec_sr}

The slowly rotating configurations of an isothermal gas (possibly
degenerate) can be determined semi-analytically by using perturbative
expansions of the Boltzmann-Poisson or Fermi-Poisson equations in
terms of the dimensionless rotation parameter $v=\Omega^{2}/2\pi
G\rho_{0}$ (Chavanis 2002c). To lowest order in the expansion, the
effect of rotation is just to flatten the system and to shift the
onset of instability (see Sec. \ref{sec_gc}). For $T=0$ (or
$E=E_{min}$), we recover the distorted polytrope of index $n=3/2$
studied by Milne (1923) and Chandrasekhar (1933). In the region of
negative specific heats, this ``rotating fermion ball'' is surrounded
by a gaseous halo. For high energies, the system is non
degenerate. Hence, the description of the equilibrium phase
diagram is very similar to that of the non-rotating Fermi gas.

\subsection{Spheroidal structures}
\label{sec_sd}

For higher values of rotation, the flattening increases and the system
takes a spheroidal structure. For $E\rightarrow E_{min}$, we have a
rapidly rotating ``fermion spheroid'' (Fig.~\ref{density}b). This
structure is completely degenerate and highly concentrated. In
particular, it is insensitive to the confining box. As energy
increases further, the spheroid is only partially degenerate and
extends to larger and larger distances (Fig.~\ref{density}e). The
rotation affects in priority the low density external region while the
dense core of the configuration remains almost spherical. In the region of
negative specific heats ($E_{cond}< E< E_{gas}$), the ``spheroid'' is
surrounded by a diffuse halo. This corresponds to a mixed phase. This
``spheroid-halo'' structure generalizes the ``core-halo'' structure of
non-rotating systems. For even larger values of energy, the spheroid
would extend outside the box so that the system becomes almost
one-dimensional and depends only on the axial coordinate $z$
(Fig.~\ref{density}h). Note, however, that the density contrast is
extremely low (${\cal R}=\rho(0)/\rho(R)\simeq 1.00022$) so that the
structure is essentially homogeneous.

\subsection{Keplerian limit}
\label{sec_k}

For $E=E_{min}$, the temperature drops to zero and the system is
equivalent to a pure polytrope with index $n=3/2$. The structure of
rapidly rotating polytropes is well-known (James 1964). For rigid
polytropes with index $n<0.808$, the system is sensitive to
non-axisymmetric instabilities of various forms. Therefore, at
sufficiently high rotations, the spheroidal sequence bifurcates
continuously to more complicated structures (Chandrasekhar 1969).
However, for $n>0.808$, non-axisymmetric instabilities are inhibited
and the series of equilibria has a different evolution. The polytrope
develops a {\it cusp} at the equator at a point where the
gravitational force balances the centrifugal force
(Fig.~\ref{density}c). This is the so-called Keplerian (or break-up)
limit corresponding to a critical angular velocity $\Omega_{max}$ such
that $\Omega_{max}^{2}R_{e}^{3}=GM$, where $R_{e}$ is the equatorial
radius.  There is no equilibrium state for $\Omega>\Omega_{max}$ as
mass is ejected from the equatorial surface by the centrifugal force
(mass shedding). The tendency to expulse some mass far away may be
related to the formation of a Keplerian disk above $\Omega_{max}$. In
that case, we must allow for differential rotation.  Mathematically,
above $\Omega_{max}$, the maximization problem leading to
Eq. (\ref{mf9}) has no solution that are everywhere
differentiable. The formation of an equatorial cusp at
$\Omega=\Omega_{max}$ marks the break-up of differentiability and the
complete change of structure. For $\Omega>\Omega_{max}$, the system
may resemble a ``cuspy condensate'' $+$ some matter accumulating at the
box (ejected by the central body). We can also wonder whether
non-axisymmetric structures can form above $\Omega_{max}$. Indeed,
there exists ``double-cluster'' solutions for polytropes with index
$n>0.808$ (Hachisu \& Eriguchi 1984) that bifurcate {\it
discontinuously} from the spheroidal sequence. On general grounds, it
would be of interest to compare the entropy of the single cluster to
the entropy of the double cluster. This problem is left for future
investigations.

For $E=E_{min}$ and $\Omega=\Omega_{max}$, all the mass is in the
cuspy polytrope. As $E$ increases along the line of maximum rotation, the
polytrope extends in size and is embedded in a disk-like halo
(Fig.~\ref{density}f).  The maximum angular velocity $\Omega_{max}(E)$
decreases with $E$. This is because the size of the polytrope
increases so that it is more easily affected by rotation.  At high
energies, the polytrope is very extended and the cusp would form
outside the confining box. Therefore, the structure of the system is
highly affected by the wall and this makes possible to achieve
arbitrary large values of angular velocity. As a result, the maximum
angular velocity increases with the energy when $E$ is large. For high
rotations and high energies, the matter accumulates on the wall due to
the centrifugal force (Fig.~\ref{density}i). The density profile is
close to the law
\begin{equation}
\rho=Ae^{{1\over 2}\beta \Omega^{2}s^{2}},
\label{k1}
\end{equation}
which corresponds to the density of a rotating gas without
self-gravity. Note that the iso-density lines are axial so that a
complete change of symmetry has occurred from Fig.~\ref{density}h to
Fig.~\ref{density}i.

\subsection{Equilibrium phase diagram}
\label{sec_pd}

\begin{figure}[htbp]
\centerline{
\includegraphics[width=8cm,angle=0]{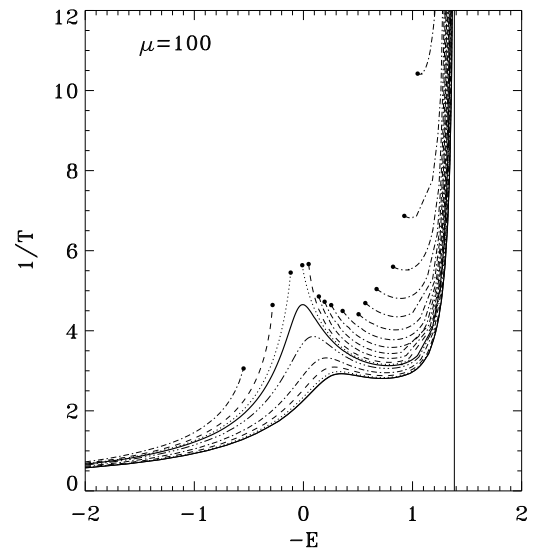}
}
\caption[]{The caloric curve of a rotating system with $\mu=100$. It has a $N$-shape structure characteristic of self-gravitating systems with large cut-off ($\mu\le \mu_{MTP}=2670$). The ``gaps'' in energy and temperature at high rotations are due to the absence of uniformly rotating solution above the
Keplerian limit. The thin full
line shows the $\Omega=0$ case while the thick line marks the entry in the Kepler zone (shaded region in Figs.  \ref{phase_micro} and \ref{phase_can}). In between lines show some
intermediate cases.}
\label{calor100}
\end{figure}

The caloric curve $T(E)$ is plotted in Fig.~\ref{calor100} for
different values of angular velocity $\Omega$. It has the typical
``$N$-shape'' structure previously found for non-rotating and slowly
rotating systems with high cut-off.  The onset of instability can be
determined by a turning point criterion (Katz 1978). For $\mu=100$,
the equilibrium states are always stable in the microcanonical
ensemble (they are maxima of entropy at fixed mass, energy and angular
momentum) but they are unstable in the canonical ensemble in the
region of negative specific heats (they are saddle points of free
energy at fixed mass, temperature and angular velocity). We see that,
for moderate rotations, the structure of the caloric curve does not
change dramatically with respect to the non-rotating case. It is just
a natural continuation of previously reported results. However, at
high rotations, the system is sensitive to mass shedding and there
exists a maximum angular velocity $\Omega_{max}$ for each value of
energy and temperature (Keplerian limit).

\begin{figure}[htbp]
\centerline{
\includegraphics[width=8cm,angle=0]{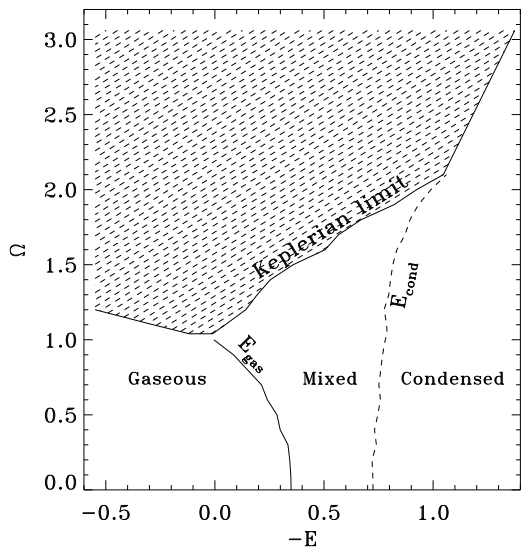}
}
\caption[]{Microcanonical phase diagram in the $(E,\Omega)$ plane
for $\mu=100$. The maximum angular velocity $\Omega_{max}(E)$ depends
on the energy. It is maximum for the $T=0$ configuration (which
corresponds to a $n=3/2$ polytrope). The
parameters of this terminal point are $\Omega_{max}=3.06$ and
$E_{min}=-1.37$.}
\label{phase_micro}
\end{figure}

The phase diagram in the $(E,\Omega)$ plane is plotted in
Fig.~\ref{phase_micro}. Since the energy is assumed to be fixed, this
diagram corresponds to the microcanonical description
\footnote{We have used the angular velocity as a
control parameter instead of the angular momentum for commodity but
also because it is the most natural variable to characterize a
rotating system. Indeed, the conservation of angular momentum should
not be taken in a strict sense because it can always be satisfied by
ejecting a small amount of mass far away with weak influence on the
other constraints. For similar remarks concerning the role of angular
momentum in the context of vortex dynamics, see Brands {et al.}
(1999).}. In continuity with the non-rotating case, the horizontal
structure of the diagram consists of three regions: a pure
``condensed'' phase for $E<E_{cond}$, a pure ``gaseous''
phase for $E>E_{gas}$ and a mixed ``core-halo'' phase for
intermediate energies. Vertically, the diagram proceeds with
increasing flattening until the Keplerian limit is reached for
$\Omega=\Omega_{max}(E)$. Above this line, the system is expected to
expulse matter far away and, possibly, create a Keplerian disk or
break into fragments.

\begin{figure}[htbp]
\centerline{
\includegraphics[width=8cm,angle=0]{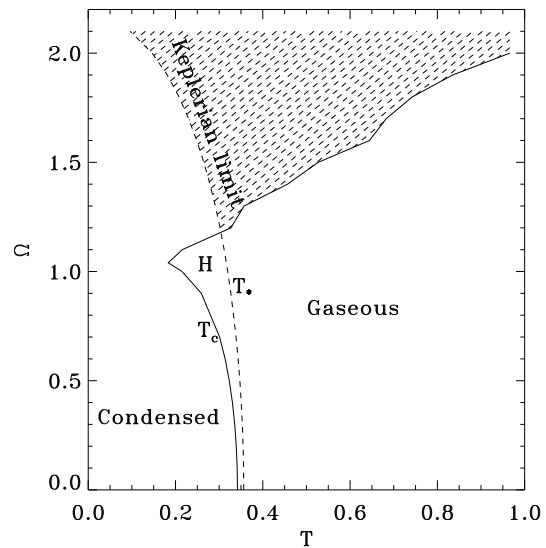}
}
\caption[]{Canonical phase diagram in the $(T,\Omega)$ plane for $\mu=100$.
The $H$-zone corresponds to an hysteretic zone where the actual phase
depends on the history of the system. }
\label{phase_can}
\end{figure}

The phase diagram in the $(T,\Omega)$ plane is plotted in
Fig.~\ref{phase_can}. Since the temperature is assumed to be fixed,
this diagram corresponds to the canonical description. The mixed phase
with negative specific heats allowed in the microcanonical ensemble is
replaced by a phase transition connecting the ``gaseous'' phase to the
``condensed'' phase. Between $T_{c}$ and $T_{*}$ the system can be in
a metastable state. If it is initially prepared in a gaseous state, it
will remain gaseous until the minimum temperature $T_{c}$ at which it
will collapse and become condensed. Inversely, if the system is
initially prepared in a condensed state, it will remain condensed
until the maximum temperature $T_{*}$ at which it will explode and
become gaseous (see Fig. \ref{calor0}). Therefore, the configuration
of the system in the $H$-zone depends on its history. This is like a
hysteretic cycle in magnetism.

Let us comment again on the distinction between ${\bf L}$ and ${\bf
\Omega}$. If we restrict ourselves to single-cluster 
axisymmetric solutions, as we do, we do not expect qualitative
difference in the phase diagram according as we fix angular momentum
or angular velocity. Indeed, there is a one-to-one correspondance
between ${\bf \Omega}$ and ${\bf L}$.  This is very different if we
allow for non-axisymmetric, or disconnected, solutions (such as double
clusters, rings,...) that are beyond the scope of the present
paper. In that case, there is no one-to-one correspondance between
${\bf \Omega}$ and ${\bf L}$ and the choice of the control parameter
is crucial. The thermal energy $K_{th}=E-W-{1\over 2}I\Omega^{2}$ is
for a single central cluster at the same $E$, $\Omega$ much greater
than for a double cluster with larger $I$ and $W$. Thus, in
$(E,\Omega)$ ensemble, the central system has more entropy (thermal
random motion) than the double cluster. Hence, the single cluster will
overshadow the probability of the binary state (we thank the referee for
this comment).  On the other hand, a double cluster rotates more
slowly than a single central cluster with same $L$. Therefore, in
$(E,L)$ ensemble, a single cluster could avoid forming an equatorial
cusp by splitting in two pieces. Depending on $L$ and $E$, the double
cluster solution may have more entropy than the single one. It would
be of interest to consider this problem in detail in the case of
self-gravitating fermions. We do not know if ``binary fermion balls''
have been introduced in dark matter models. We can wonder whether the
transition from single to double clusters (if any) occurs at or before
the Keplerian limit. Again, we emphasize that in the case of
self-gravitating fermions (similar to polytropes $n=3/2>0.808$), this
transition would be discontinuous with respect to the spheroidal
sequence.

\section{Caloric curves for high values of the degeneracy parameter 
(low cut-off/large system) }
\label{sec_cc}

We now consider the case of a high value of the degeneracy parameter
$\mu=10^{5}>\mu_{MTP}=2670$ (large system) so that the influence of
the small-scale cut-off is weaker in average and the conditions
required to set up a gravothermal catastrophe are fulfilled.

\begin{figure}
\centerline{
\includegraphics[width=8cm,angle=0]{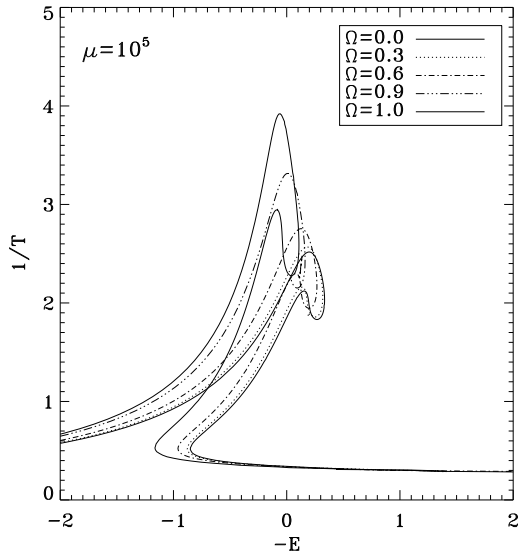}
}
\caption[]{The caloric curve of a rotating system with $\mu=10^{5}$. It has a $Z$-shape structure (dinosaur's neck) characteristic of self-gravitating systems with small cut-off ($\mu\ge\mu_{MTP}=2670$). The ground state ($T=0, E=-138$) is outside the range of represented energies.}
\label{calor5}
\end{figure}

\begin{figure}
\centerline{
\includegraphics[width=8cm,angle=0]{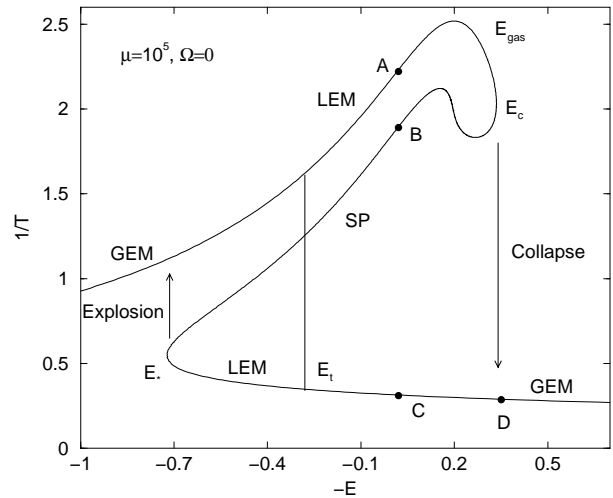}
}
\caption[]{Caloric curve of a self-gravitating system with $\mu=10^{5}$ (here $\Omega=0$). Between $E_{t}$ and $E_{c}$ (resp. $E_{*}$) the gaseous states (resp. condensed states) are {\it metastable}. A collapse occurs at
$E_{c}$ (gravothermal catastrophe) and an explosion occurs at
$E_{*}$. By varying the energy of the system between $E_*$ and $E_c$
(spinodal points), we can generate an hysteretic cycle in the
microcanonical ensemble. }
\label{hystmicro}
\end{figure}

The caloric curve $T(E)$ is plotted in Fig.~\ref{calor5} for different
values of angular velocity $\Omega$. It has the typical ``$Z$-shape''
structure (resembling a ``dinosaur's neck'') previously found
\footnote{Of course, in a strict sense, the microcanonical caloric
curve must be single valued everywhere because $S_{micro}(E)=\ln g(E)$
and $1/T_{micro}(E)=dS_{micro}/dE$ (Gross 2001,2003). However, the
caloric curves that we represent are richer because they contain
saddle points SP (unstable), local maxima LEM (metastable) and global maxima
GEM (stable) of entropy. Had we represented only global entropy maxima,
the curves $S(E)$ and $\beta(E)$ would be single-valued. They would
describe the true statistical equilibrium states reached for
$t\rightarrow +\infty$. However, metastable states are of considerable
importance in astrophysics because they correspond to the observed
structures (e.g., globular clusters) for the timescales contemplated
(as is well known, stellar systems are {\it not} in true statistical
equilibrium states). Therefore, the physical caloric curves must take
these metastable states into account and this gives rise to what we
have nicknamed ``dinosaur's necks''. We recall that these metastable
states appear only for sufficiently small cut-offs. For large cut-offs
(see Fig. \ref{calor0}), there is only one global entropy maximum for
each energy and the $\beta(E)$ curve is univalued. This is the 
situation considered by Votyakov {\it et al.} (2002) [compare their
Fig. 7 to our Fig. 3].} for non-rotating or slowly rotating systems
(Chavanis \& Sommeria 1998, Chavanis 2002b,c). The upper branch
corresponds to non-degenerate states (gaseous phase) and the lower
branch corresponds to core-halo states (condensed phase). A first
order {\it microcanonical} phase transition is expected to occur at
$E_{t}(\mu)$ since at this energy the two phases have the same entropy
(Fig.~\ref{hystmicro}). This gravitational phase transition is marked
by a discontinuity of temperature and specific heats.  If
$E_{t}>E_{gas}$ (where $E_{gas}$ is the first turning point of
temperature), the specific heats passes from positive to negative
values. If $E_{t}<E_{gas}$, the specific heats is always negative at
the transition (the crossing point occurs for $\mu\simeq 1.10\ 10^4$).
However, this first order phase transition may not take place in
practice because the gaseous states with energy $E_{c}<E<E_{t}$ are
metastable (local entropy maxima) and long-lived (Katz \& Okamoto
2000, Chavanis
\& Ispolatov 2002).  The collapse (gravothermal catastrophe) will set
in at, or near, the critical energy $E_{c}$ (Antonov energy) at which
the gaseous phase disapears. This corresponds to a zeroth order
microcanonical phase transition accompanied by a discontinuity of
temperature and entropy. The collapse stops when the central region
becomes degenerate and forms a dense nucleus. The resulting
configuration has a ``core-halo'' structure. Contrary to the collapse
in the canonical ensemble, the nucleus contains only a {\it moderate}
fraction of mass. In the classical limit $\mu\rightarrow +\infty$, it
reduces to a single binary (see Sec. \ref{sec_gc}). The rest of the
mass is diluted in a hot envelop (with almost uniform density) that is
held by the walls of the box. For an open system, the halo would be
dispersed at infinity so that only the degenerate nucleus would
remain. This behaviour, with the expulsion of an envelope, is more
consistent with the process of white dwarf formation than a complete
collapse of the system at fixed $T$. This is an astrophysical evidence
that the microcanonical ensemble is more appropriate than the
canonical one.  The condensed states with energy $E_{t}<E<E_{*}$ are
also metastable and long-lived. An ``explosion'', reversed to the
gravothermal catastrophe, will occur above $E_*$ if the system is
initially prepared in the condensed phase (see
Fig.~\ref{hystmicro}). By varying the energy between $E_*$ and $E_c$
(microcanonical spinodal points) we can generate an {\it hysteretic
cycle} in the microcanonical ensemble (Chavanis 2003b).

\begin{figure}
\centerline{
\includegraphics[width=8cm,angle=0]{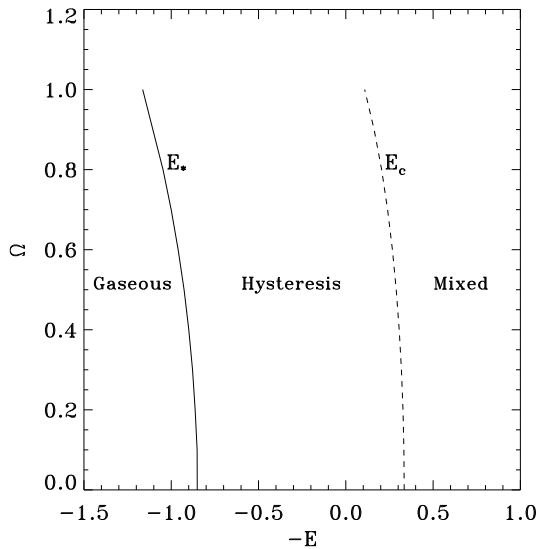}
}
\caption[]{Microcanonical phase diagram for $\mu=10^5$. The $H$-zone corresponds to an hysteretic zone where the actual phase depends on the history of the system. }
\label{phase5}
\end{figure}

The phase diagram describing the microcanonical phase transition is
represented in Fig.~\ref{phase5}. It can be compared with the phase
diagram describing the canonical phase transition for $\mu=100$ (see
Fig.~\ref{phase_can}), the role of energy $E$ and temperature $T$
being reversed. In particular, there exists a $H$-zone in which the
structure of the system (gas or condenstate) depends on its
history. For smaller energies ($E<E_{c}$), the system is in a mixed
phase with negative specific heats. For even lower energies
$E_{cond}\sim -117$ (not represented on the diagram), we leave the
region of negative specific heats and enter the pure condensed phase
as in Fig.~\ref{phase_micro}.  More details on these phase transitions
can be found in Chavanis (2002b). It is shown in particular that the
microcanonical first order phase transition disapears for
$\mu<\mu_{MTP}=2670$.

\section{Gravothemal catastrophe for classical self-gravitating particles (no cut-off)}
\label{sec_gc}

We now consider the limit $\mu= +\infty$ (or $h=0$) corresponding to
classical particles. In that limit, the particles do not feel any kind
of exclusion or small-scale cut-off. This is the situation relevant
for globular clusters for which the size of the stars does not matter.

\begin{figure}
\centerline{
\includegraphics[width=8cm,angle=0]{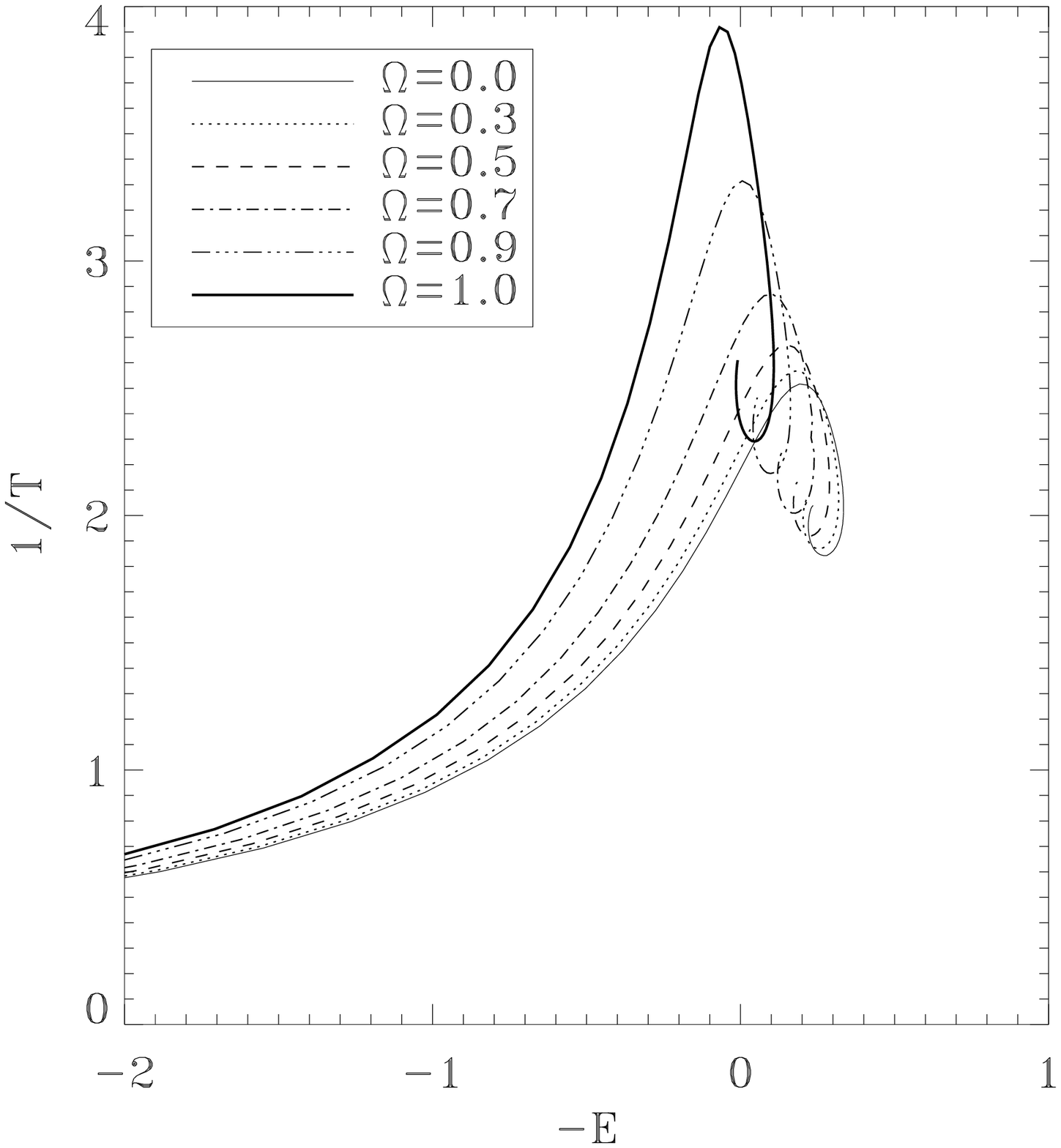}
}
\caption[]{Caloric curves corresponding to classical point masses ($\mu=+\infty$) for different values of angular velocity.
Below $E_{c}$, the gravothermal catastrophe leads to the formation of
a binary star (microcanonical ensemble). Below $T_{c}$, the isothermal
collapse generates a Dirac peak (canonical ensemble).  }
\label{calor}
\end{figure}

\begin{figure*}
\centerline{
\includegraphics[width=9cm,angle=90]{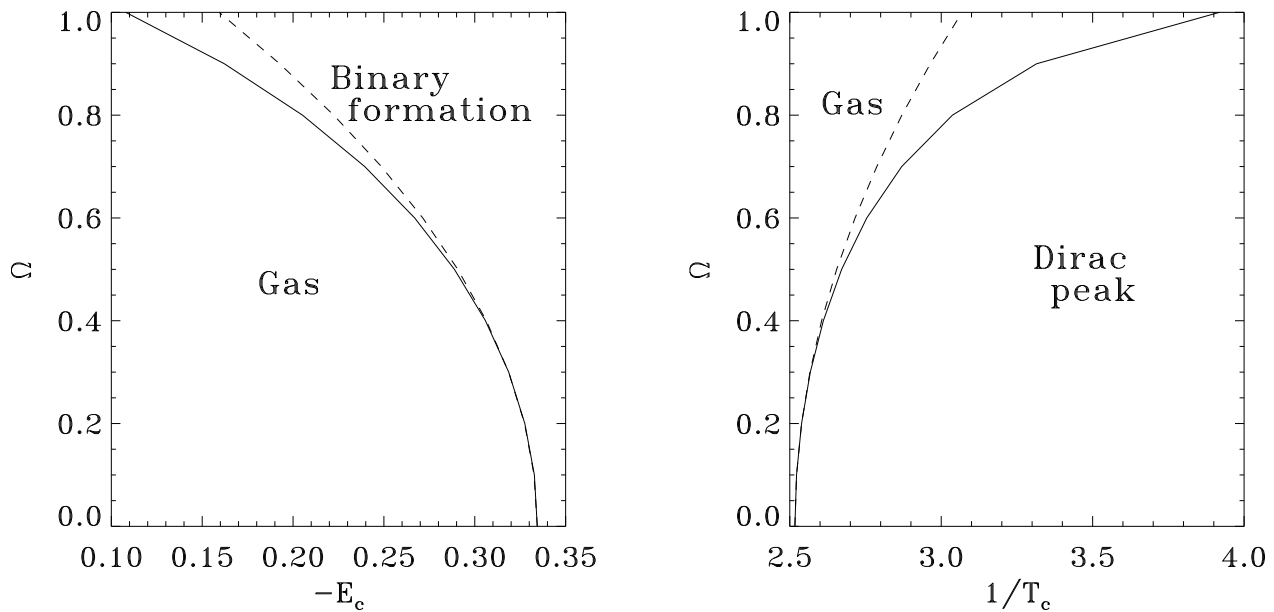}
}
\caption[]{Dependence of the Antonov critical energy and critical
temperature with rotation. For high energies or high temperatures, the
system is in a gaseous (metastable) phase. Below a critical energy or
temperature, it enters in the collapsed phase. In the microcanonical
ensemble (fixed $E$), the collapse leads to a single binary (or a
small group of stars) surrounded by a hot dilute halo. In the
canonical ensemble (fixed $T$), the collapse leads to a Dirac peak
containing all the mass (``black hole''). This Dirac peak forms in the {\it post-collapse}
regime of the dynamical evolution. For small rotations, the critical
energy and critical temperature behave as
$E_{c}(\Omega)=E_{c}(0)+0.175\Omega^{2}$ and
$1/T_{c}(\Omega)=1/T_{c}(0)+0.552\Omega^{2}$ (dotted lines). This
scaling is valid typically for $\Omega\le 0.5$. }
\label{antonov}
\end{figure*}

In the non degenerate limit, the caloric curve has a classical spiral
shape (Fig.~\ref{calor}). We recall that, in the absence of
small-scale cut-off, the equilibrium states are only metastable as
they correspond to {\it local} maxima of the thermodynamical potential
(entropy or free energy). These metastable states play however an
important role as they correspond to the observed structures (see
discussion in Chavanis 2003a). In the microcanonical ensemble, the
gravothermal catastrophe sets in at the Antonov energy $E_{c}$.  The
microcanonical phase diagram in $(E,\Omega)$ plane is drawn in
Fig.~\ref{antonov} and shows in particular the evolution of the
Antonov critical energy $E_{c}(\Omega)$ with the rotation.  The
critical energy $E_{c}(\Omega)$ increases with rotation so that
gravitational collapse occurs {\it sooner} than in the non-rotating
case. This confirms the results of the perturbative approach (Chavanis
2002c). Qualitatively, the critical energy increases due to the
rotational term ${1\over 2}I\Omega^{2}>0$ in the expression of $E$ (see
Sec. \ref{sec_ic}). The result of the collapse is to form a {\it
binary} star with large binding energy. Since total energy is
conserved, the gravitational energy released by the binary is
redistributed in a dilute halo in the form of thermal energy. This
evolution is natural in the microcanonical ensemble since it can
produce very high (ideally diverging) values of entropy due to the
divergence of the temperature. It is also vindicated by numerical
simulations of the orbit averaged Fokker-Planck equation (Cohn 1980)
or fluid equations (Lynden-Bell \& Eggleton 1980). These
simulations show that the collapse of globular clusters is
self-similar and leads to a finite time singularity with a density
profile $\rho\sim r^{-\alpha}$ with $\alpha\simeq 2.2$. The mass
contained in the core goes to zero (indicating that the collapse is
halted by the formation of binaries with mass $2m\ll M$) while the
central temperature raises to $+\infty$ since $\alpha>2$. Binary-like
structures are also found in the microcanonical model of
self-gravitating Brownian particles (Sire \& Chavanis 2003).

In the canonical ensemble, the collapse sets in at the critical
temperature $T_{c}$. The evolution of this critical temperature with
the angular velocity is represented in Fig.~\ref{antonov}. It is found
that $T_{c}(\Omega)$ {decreases} with $\Omega$ so that gravitational
collapse is {\it delayed} in the canonical ensemble. This corroborates
the results found with the perturbative approach (Chavanis 2002c). The
result of the collapse is to form a {\it Dirac peak}, i.e. a state
where all the mass is concentrated at the center (``black hole'').
This evolution is natural in the canonical ensemble since it can
produce very high (in fact diverging) values of free energy $J=S-\beta
E$ due to the divergence of the potential energy. This isothermal
collapse can be followed dynamically by solving the Euler-Jeans equations
for a gaseous system with an equation of state $p=\rho T$ where $T$ is
fixed (Penston 1969). It can also be studied analytically in the canonical
model of self-gravitating Brownian particles (Chavanis {et al.}
2002). In this model, there is first a finite time singularity leading
to a $r^{-2}$ density profile. Then, the collapse continues after the
singularity is formed and the Dirac peak appears in the post-collapse
regime due to the accretion of mass (Sire
\& Chavanis 2002,2003). We expect this two-stage behaviour to persist in
the case of more relevant dynamical models.  This may be an important
point for numerical simulations in astrophysics and cosmology because
if we only consider the first regime we conclude that the core
contains no mass (while its density is high and scales as
$r^{-2}$). However, in a post-collapse regime, a Dirac peak is formed
which contains a macroscopic mass at a single point. We can wonder
whether this mechanism (apparently not reported before) is related to
the existence of black holes at the center of galaxies. In reality,
small-scale cut-offs must be introduced (e.g., quantum degeneracy) and
the black hole singularity is replaced by a ``fermion ball'' with a
high mass. Note that in the microcanonical ensemble, there is no black
hole singularity for classical point masses (the Dirac peak is
replaced by a binary). Interestingly, for self-gravitating fermions,
the condensate (fermion ball) contains a moderate but {\it
macroscopic} fraction of mass (Chavanis
\& Sommeria 1998). This may explain the formation of massive objects
at the center of galaxies in a microcanonical framework. The
microcanonical description is the most relevant as it accounts for the
transfers of energy between the core and the halo contrary to isothermal
models.

The divergence of entropy in the microcanonical ensemble due to the
formation of a binary embedded in a hot halo is implicit in the work
of Antonov (1962). It is also clear from the arguments given by
Padmanabhan (1990) to show that the density of state diverges as we
approach two particles to each other provided that thermal energy can
be redistributed in a halo (this demands $N\ge 3$). The divergence of
free energy in the canonical ensemble by forming a Dirac peak has been
shown rigorously by Kiessling (1989) and heuristically by Chavanis
(2002e). The inequivalence of statistical ensembles regarding the
formation of binaries or Dirac peaks is further discussed in
Appendices A and B of Sire \& Chavanis (2002). In this approach, a
family of distributions functions is constructed which provokes the
divergence of entropy or free energy when a physical parameter is
varied. These results can also be obtained by considering the small
cut-off limit $\mu\rightarrow +\infty$ in the analytical model of
phase transitions for self-gravitating systems proposed by Chavanis
(2002b).

\section{Phase diagram of non-rotating systems}
\label{sec_pdnr}

In previous sections, we have presented the equilibrium phase diagram
of self-gravitating fermions as a function of  angular velocity
$\Omega$ for a fixed value of degeneracy parameter $\mu$. We shall now discuss
the equilibrium phase diagram of self-gravitating fermions (or hard
sphere systems) as a function of $\mu$ for $\Omega=0$. These phase diagrams
have not been published before.

\begin{figure}
\centerline{
\includegraphics[width=8cm,angle=0]{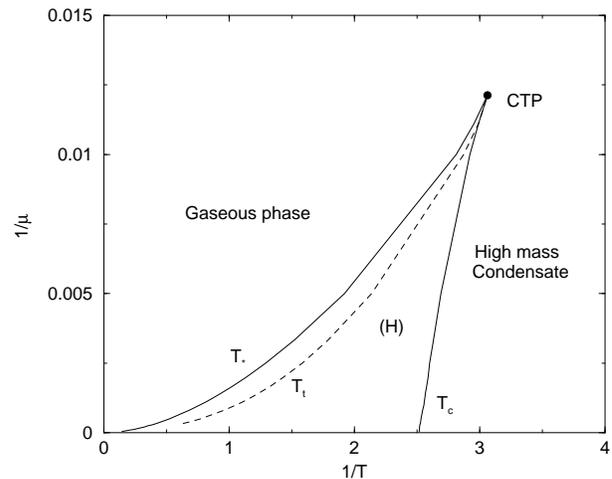}
}
\caption[]{Canonical phase diagram in $(T,\mu)$ plane for $\Omega=0$. The $H$-zone corresponds to an hysteretic zone where the actual phase depends on the history of the system. }
\label{phaseTP}
\end{figure}

The deformation of the caloric curve $T(E)$ as a function of the
degeneracy parameter $\mu$ is described in detail by Chavanis
(2002b). This work completes earlier studies on phase transitions in
self-gravitating systems by showing the small cut-off limit,
characterized by an unwound spiral (``dinosaur's neck''), leading to a
first order microcanonical phase transition (earlier works essentially
described the canonical first order phase transition).  Typical curves
illustrating first order canonical and first order microcanonical
phase transitions are shown in Figs. \ref{calor0} and
\ref{hystmicro}. The equilibrium phase diagram of self-gravitating
fermions can be directly deduced from these results by identifying
characteristic energies and temperatures. In the canonical ensemble,
we note $T_{t}$ the temperature of transition (determined by the
equality of the free energies of the two phases), $T_{c}$ the end
point of the metastable gaseous phase (first turning point of
temperature) and $T_{*}$ the end point of the metastable condensed
phase (last turning point of temperature). The temperature of
transition $T_{t}$ is determined by a horizontal Maxwell construction
(Chavanis 2003b). The canonical phase diagram is shown in
Fig. \ref{phaseTP}. It shows in particular the canonical critical
point $\mu_{CTP}=83$ at which the canonical first order phase
transition disappears and is replaced by a localized second order
phase transition.

\begin{figure}
\centerline{
\includegraphics[width=8cm,angle=0]{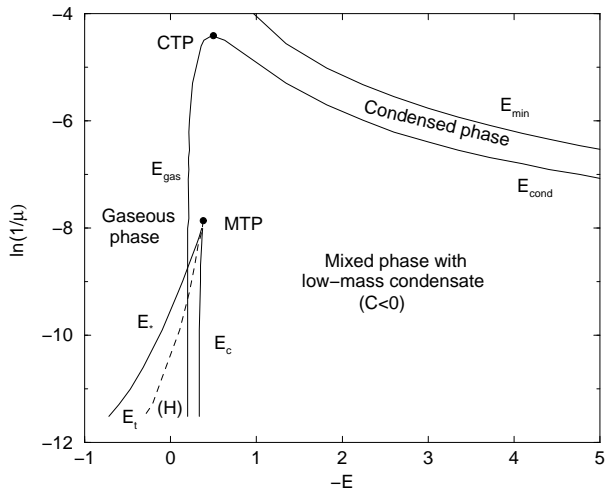}
}
\caption[]{Microcanonical phase diagram in $(E,\mu)$ plane for $\Omega=0$.  The phase diagram in MCE is more complex than in CE due to the existence of the negative specific heat region that is forbidden in CE. }
\label{phaseEP}
\end{figure}

In the microcanonical ensemble, we note $E_{t}$ the energy of
transition (determined by the equality of the entropy of the two
phases), $E_{c}$ (Antonov energy) the end point of the metastable
gaseous phase (first turning point of energy) and $E_{*}$ the end
point of the metastable condensed phase (last turning point of
energy). The energy of transition $E_{t}$ is determined by a less
standard vertical Maxwell construction (Chavanis 2003b). We also
denote by $E_{gas}$ the energy at which we enter in the mixed phase
with negative specific heat (first turning point of temperature) and
$E_{cond}$ the energy at which we leave the mixed phase and enter the
condensed phase (last turning point of temperature). We also introduce
the minimum energy $E_{min}$ (ground state). The microcanonical phase
diagram is shown in Fig. \ref{phaseEP}. It shows in particular the
microcanonical critical point $\mu_{MTP}=2670$ at which the
microcanonical first order phase transition disappears. The structure
of the equilibrium phase diagrams can be easily understood in the
light of the preceding discussion (see Chavanis 2002b for more
details). We note that the microcanonical phase diagram is more
complex than the canonical one due to the existence of a negative
specific heat region.  We recall that canonical stability (maximum of
$J$ at fixed $M$, $\beta$) is a sufficient, albeit not necessary,
condition of microcanonical stability (maximum of $S$ at fixed $M$,
$E$). Since canonical equilibria are always realized as microcanonical
equilibria, they constitute a sub-domain of the microcanonical phase
diagram.  Hence, the microcanonical ensemble is richer than the
canonical one, as is well-known.

\section{Conclusion}
\label{sec_conc}

We have calculated single-cluster axisymmetric statistical equilibrium
states of self-gravitating fermions for arbitrary values of rotation
and temperature (or energy). The problems associated with the
divergence of the gravitational potential as $r\rightarrow 0$ have
been avoided by considering quantum effects. This small-scale
regularization is a physical one because the Pauli exclusion principle
is a fundamental concept and fermionic matter is relevant in
astrophysics.  Isothermal and polytropic configurations (with index
$n=3/2$) are obtained as particular limits of our model ($T\rightarrow
+\infty$ and $T=0$).  On the other hand, depending on the importance
of rotation, we have obtained different types of structures such as
fermion balls, fermion spheroids and cuspy structures. In the mixed
phase, these condensed objects are surrounded by a diffuse isothermal
halo.  Clearly, the main drawback of our study is the necessity to
enclose the system within an artificial box so as to prevent its
complete evaporation (which is the true statistical equilibrium state
for self-gravitating systems). A future extension of our study is to
consider more realistic truncated models in order to avoid the
artifice of a material box and take into account incomplete relaxation
or tidal effects. We note, in passing, that the distribution function
$f=f(\epsilon_{J})$ predicted by statistical mechanics cannot account
for the triaxial structure of elliptical galaxies since no bifurcation
from axisymmetric (spheroidal) to non-axisymmetric (ellipsoidal)
structures are expected for uniformly rotating compressible
bodies. The distribution function must depend on isolated integrals
different from the single Jacobi integral (Contopoulos 1960). This is
clearly a manifestation of incomplete relaxation but the construction
of such isolated integrals is a highly complicated task and demands a
very delicate analysis of the process of formation of elliptical
galaxies.

We emphasize that the nature of the small-scale cut-off is of
considerable importance for rotating objects, regarding the structure
of the condensed phase, because it can generate or inhibit
non-axisymmetric instabilities.  For example, the ordinary Fermi-Dirac
distribution used in this paper and the model of Votyakov et
al. (2002) have the same high temperature limit (an isothermal gas
with $p=\rho T$) but a different low temperature limit. The Fermi gas
at $T=0$ corresponds to the equation of state $p=K\rho^{5/3}$
(polytropic core).  The equation of state corresponding to a
Fermi-Dirac distribution in configuration space
$\rho=\rho_{max}/\lbrack 1+\lambda {\rm exp}(\beta\Phi_{eff})\rbrack$
is $p=T\ln(1-\rho/\rho_{max})$, as can be obtained from the condition
of hydrostatic equilibrium \footnote{A Fermi-Dirac distribution in
configuration space also arises in the statistical mechanics of
two-dimensional vortices (see Chavanis 2002a for a review on the
analogy between stellar systems and 2D vortices). In that respect, it
may be relevant to mention the work of Chen \& Cross (1996) who found
``double-vortex'' equilibrium solutions in a circular domain when the
conservation of angular momentum is accounted for. This can be viewed
as the 2D hydrodynamical version of the ``double-star'' structure
found by Votyakov et al. (2002). Note that the presence of a confining
box is crucial to maintain the double-vortex solution.  Similarly, for
self-gravitating systems at non-zero temperature, an artificial box is
needed to maintain the double-star structure.}. For $T=0$, $\rho$ is a
step function with $\rho\rightarrow
\rho_{max}$ for $r<R_{*}$ (solid core) and $\rho=0$ otherwise. In the
former case (fermions), there is no bifurcation to non-axisymmetric
structures at high rotations  in continuity with the
spheroidal sequence. Instead, the system develops an equatorial cusp
(the possibility of {discontinuous} bifurcations to
non-axisymmetric structures will be considered in a future work). This
cusp does not exist for homogeneous bodies (or hard sphere systems)
and non-axisymmetric bifurcations occur continuously at high
rotations. It is surprising that the equation of state corresponding
to the model of Votyakov et al. (2002) does not coincide with the Van
der Waals equation of state, as in the work of Aronson
\& Hansen (1972), since these authors treat their
system as a hard sphere gas \footnote{In fact, Votyakov et al (2002)
prevent overlapping of particles in physical space, as in a lattice
gas model. This is a simple way to limit the local density $\rho({\bf
r})$ of the self-gravitating gas. Hence, their model is expected to
represent stars at densities up to the ignition of their hydrogen
burning, when further gravitational collapse is halted for a while
(Gross 2003). In this context, it may be relevant to consider a large
small-scale cut-off, as they do. Note that their model does not
correspond to the Lynden-Bell (1967) statistics, which is a
Fermi-Dirac distribution in phase space appropriate to collisionless
stellar systems (e.g., Chavanis 2002d).}. It is also surprising that
they do not obtain a richer variety of structures in the low
temperature/energy regime (pear shaped, dumbbell,...) since rotating
homogeneous bodies possess a rich bifurcation diagram as investigated
since the $19$-th century (Chandrasekhar 1969).  These remarks show
that the small-scale regularization must be given precise
consideration depending on the situation contemplated.

The self-gravitating Fermi gas has several astrophysical applications:
(i) Lynden-Bell's type of degeneracy may be relevant for galactic
nuclei and dark matter (Lynden-Bell 1967, Kull et al. 1996, Chavanis
\& Sommeria 1998, Stiavelli 1998, Leeuwin \& Athanassoula 2000).  
(ii) Pauli's exclusion principle is relevant for white dwarfs and
neutron stars (Chandrasekhar 1942, Hertel \& Thirring 1971).  In that
context, the Fermi gas at non-zero temperature provides a simple
theoretical model describing a phase transition from a ``gaseous
star'' (for $T>T_{c}$) to a white dwarf or a neutron star (for
$T<T_{c}$). To obtain a better description of these objects, special
relativity must be taken into account for massive white dwarf stars
(Chandrasekhar 1931) and general relativity is required in the case of
neutron stars (Oppenheimer \& Volkoff 1939).  (iii) It has been
proposed recently that {dark matter} could be made of a collisionless
gas of massive neutrinos. Therefore, the relevant distribution
function is again the Fermi-Dirac distribution (see, e.g., Chavanis
2002f). The degeneracy is due either to the Liouville theorem (in the
context of violent relaxation) or to Pauli's exclusion principle (if
quantum effects are relevant). By cooling below a critical
temperature, this neutrino gas is expected to undergo a phase
transition to a compact object called a ``fermion ball''. It has been
proposed that fermion balls could provide an alternative to black
holes at the center of galaxies (Bilic \& Viollier 1997). (iv)
Incidentally, our study can also be connected to models of rotating
stars (e.g., Roxburgh et al.  1965). Low mass stars ($M\la 1.5\
M_{\odot}$) of the main sequence have a radiative core and a
convective envelope. Massive stars ($M>1.8\ M_{\odot}$) have the
opposite configuration. Now, the convective region corresponds to a
polytrope of index $n=3/2$. Coincidentally, this case is precisely
covered by our study.

On the other hand, hard spheres models have less direct applications
in astrophysics because the inter-particle distance is always
considerably larger than their size. Indeed, globular clusters can be
either in dilute metastable equilibrium states (Michie 1963, King
1966) or in a collapsing phase ending up with the formation of
binaries (H\'enon 1961). These binaries can release sufficient energy
to drive a re-expansion of the system (Inagaki \& Lynden-Bell 1983,
Bettwieser \& Sugimoto 1984) so that complete collapse is prevented in
practice. In each case, the size of the stars does not matter. The
size of the atoms in stars does not matter neither because radiative
transfers keep the star in a gaseous phase far from condensation. When
a star collapses (because it has burned its nuclear fuel), gravity is
balanced by quantum pressure not by the pressure of a hard sphere
gas. In that respect, it may be useful to recall that the hard sphere
model of Aronson \& Hansen (1972) was presented as a crude model for
neutron stars where the hard spheres were introduced to mimick quantum
degeneracy. The hard-sphere model can however find physical
applications in the context of planet formation resulting from the
gravitational collapse of dust particles in the solar nebula (Chavanis
2000). It has also been used by Stahl et al. (1994) to describe the
formation of stars or planetoids (considered as the solid phase)
resulting from the condensation of a molecular cloud (considered as
the gaseous phase). In this context, the explosion, reverse to the
collapse, may be related to the supernova phenomenon.

Finally, our study has a direct relevance in terms of fundamental
statistical mechanics. Indeed, the different types of phase
transitions arising in the self-gravitating Fermi gas (Chavanis 2002b)
display a very rich structure: inequivalence of statistical ensembles,
negative specific heats, convex intruder in the entropy-energy curve,
canonical and microcanonical first order phase transitions, second
order phase transition, zeroth order phase transition, canonical and
microcanonical critical points, metastable states, hysteretic
cycles,... Such properties are probably common to other systems with
long-range interactions. In fact, the self-gravitating Fermi gas forms
a generic example of systems displaying codimensions $0$ and $1$
singularities in the classification of phase transitions (for
long-range interactions) recently proposed by Bouchet \& Barr\'e
(2003).  Therefore, the detailed study of phase transitions in the
self-gravitating Fermi gas at non-zero temperature is of interest both
for statistical mechanics and astrophysics. This point is important
because the statistical mechanics of self-gravitating systems was
considered until recently, as a curiosity, not to say a fallacy, by
statistical mechanicians.

An extension of our study would be to compute more complex
configurations of rapidly rotating self-gravitating fermions and
$n=3/2$ polytropes (double clusters, rings and core-ring sequences) in
both microcanonical (fixed $E$, $M$ and ${\bf L}$) and canonical
(fixed $T$, $M$ and ${\bf\Omega}$) ensembles. These solutions are
expected to bifurcate {\it discontinuously} from the spheroidal
sequence in contrast to hard sphere systems and homogeneous bodies
($n=0$). It would be important (but difficult) to investigate the {\it
stability} of these various configurations. For rigid bodies
($n<0.808$ polytropes and hard sphere systems at low temperature), the
spheroidal sequence becomes unstable (saddle points of $S$ or $J$)
above a certain rotation at which the branch of non-axisymmetric
structures appears. For high rotations, double clusters are global maxima of
$S$ or $J$ and the ring sequence is always unstable (saddle points of
$S$ or $J$).  For compressible bodies ($n>0.808$ polytropes and
fermionic matter at low temperature), the spheroidal sequence is
expected to remain stable (maxima of $S$ or $J$) until the Keplerian
limit since no continuous bifurcations exist. However, it is possible
that, for high rotations, the single cluster structures computed in this
paper are only {\it metastable} (local maxima of $S$ or $J$) and that
the true equilibrium states (global maxima of $S$ or $J$) are double
clusters or core-ring configurations.  In case when several stable
solutions exist for the same values of $E,{\bf L}$ (or $\beta$, ${\bf
\Omega}$), it may be important to compare their entropy and free
energy. However, the selection of the physical equilibrium state will
depend more on a notion of {\it basin of attraction} (see, e.g.,
Chavanis {\it et al.} 2002 in a related context) than whether it is a
global or a local maximum of $S$ or $J$. Thus, a dynamical study is
required.  We hope to come to these problems in future communications.
The present paper is only a first step in that direction. To be more
realistic, the box could be avoided by using the distribution function
(\ref{box1}) which provides a more physical confinement of the
system. We also emphasize that in a large domain, the conservation of
angular momentum can be satisfied by ejecting a weak amount of mass at
large distances, for example in the form of a Keplerian disk. This is
the typical structure of a protoplanetary nebula. Indeed, during the
formation of a (single) star by gravitational collapse of a molecular
cloud, about $99\%$ of the intial angular momentum is spread in a
protoplanetary disk while $99\%$ of the mass remains in the star
itself whose internal structure is hardly affected by rotation.

\vskip0.2cm
\noindent{\it Acknowledgements.} We thank the referee, D.H.E. Gross, 
for a critical reading of the manuscript. Although we share a
different point of view concerning the importance of metastable states
in astrophysics, his comments were very useful (a criticism raised in
his referee report is available on cond-mat/0307535). One of us (PHC)
is also grateful to O. Fliegans, I. Ispolatov, J. Katz and V. Laliena
for valuable discussions.

\end{document}